\documentclass[useAMS,usenatbib]{mnras}

\usepackage{graphicx}
\usepackage{longtable}
\usepackage{amssymb,amsmath}
\usepackage{hyperref}
\usepackage[dvipsnames]{xcolor}
\usepackage{tikz,hyperref}
\usepackage{ulem}

\definecolor{lime}{HTML}{A6CE39}
\DeclareRobustCommand{\orcidicon}{
        \begin{tikzpicture}
        \draw[lime, fill=lime] (0,0)
        circle [radius=0.13]
        node[white] {{\fontfamily{qag}\selectfont \tiny ID}};
        \draw[white, fill=white] (-0.0625,0.095)
        circle [radius=0.007];
        \end{tikzpicture}
        \hspace{-2mm}
}

\foreach \x in {A, ..., Z}{\expandafter\xdef\csname orcid\x\endcsname{\noexpand\href{https://orcid.org/\csname orcidauthor\x\endcsname}
                        {\noexpand\orcidicon}}
}

\newcommand{\ha}{H$\alpha$}
\newcommand{\hb}{H$\beta$}
\newcommand{\ewhb}{EW(H$\beta$)}
\newcommand{\hii}{H{\sc ii}\,}
\newcommand{\msun}{$M_{\odot}$\,}

\newcommand{\Msy}{$M_{\odot}\,{\rm yr}^{-1}$\,}
\newcommand{\ecms}{ergs\,cm$^{-2}$\,s$^{-1}$\,}
\newcommand{\kms}{km\,s$^{-1}$}

\title[UVIT images of the Cartwheel ring galaxy]
{Star formation history of the post-collisional Cartwheel galaxy using Astrosat/UVIT FUV images}
\author[Y.\,D.\,Mayya et al.]{Y.\,D.\,Mayya\thanks{Email: ydm@inaoep.mx}$^{1\orcidA{}}$,
Sudhanshu Barway$^{2\orcidB{}}$, V.\,M.\,A. G\'omez-Gonz\'alez$^{3\orcidC{}}$, \and 
J. Zaragoza-Cardiel$^{1,4,5\orcidD{}}$
\\
$^{1}$Instituto Nacional de Astrof{\'\i}sica, \'Optica y Electr\'onica, Luis Enrique Erro 1, Tonantzintla 72840, Puebla, Mexico\\
$^{2}$Indian Institute of Astrophysics (IIA), II Block, Koramangala, Bengaluru 560 034, India \\
$^{3}$Institute for Physics and Astronomy, Universit\"{a}t Potsdam, Karl-Liebknecht-Str. 24/25, D-14476 Potsdam, Germany\\
$^{4}$Consejo Nacional de Humanidades, Ciencias y Tecnolog\'ias, Av. Insurgentes Sur 1582, 03940,  Mexico City, Mexico \\
$^{5}$Centro de Estudios de F\'isica del Cosmos de Arag\'on (CEFCA), Plaza San Juan, 1, 44001, Teruel, Spain.
\\
}

\begin{document}
\maketitle

\begin{abstract}
We present the results obtained by analysing new Astrosat/UVIT far ultraviolet (FUV) image of the collisional-ring galaxy Cartwheel. 
The FUV emission is principally associated with the star-forming outer ring, with no UV detection  from the nucleus and inner ring. A few sources are detected in the region between the inner and the outer rings, all of which lie along the spokes. 
The FUV fluxes from the detected  sources are combined with aperture-matched multi-band photometric data from archival images to explore the post-collision star formation history of the Cartwheel. The data were corrected for extinction using  $A_{\rm V}$ derived from the Balmer decrement 
ratios and commonly used extinction curves. We find that the ring regions contain stellar populations of wide range of ages,
with the bulk of the FUV emission coming from non-ionizing stars, formed over the last 20 to 150~Myr, that are 
$\sim$25 times more massive than the ionizing populations. On the other hand, regions belonging to the spokes have negligible current star formation, with the age of the dominant older population systematically increasing as its distance from the outer ring increases. The presence of populations of a wide range of ages in the ring suggests that the stars formed in the wave in the past were dragged along it to the current position of the ring. We derive an average steady star formation rate, SFR $=5$~\Msy, over the past 150~Myr, with an increase to $\sim$18~\Msy\ in the recent 10~Myr.
\end{abstract}


\section{Introduction}

Collisional ring galaxies are a class of star forming galaxies that form most of their stars currently along a galaxy-sized (several kiloparsecs) ring \citep{Appleton1996}. The Cartwheel galaxy is considered as an archetype for this phenomenon \citep{Marcum1992}, where the star-forming ring measures 44~kpc in diameter (ring diameter $=72$~arcsec; distance 125~Mpc using $H_0=71$ km\,s$^{-1}$\,Mpc$^{-1}$; scale: 1~arcsec $=600$~pc).
Coherent star formation (SF) over such a large scale requires a triggering mechanism. It was demonstrated by \citet{Lynds1976} that the trigger comes from an outwardly propagating density wave, which is set in when a compact galaxy plunges 
in to a large gas-rich disk galaxy, almost perpendicular to the disk with the point of 
impact close to its center. As the wave moves outward radially, it compresses the gas and triggers 
SF along a circular ring, leaving behind an aging stellar population in its wake. 

The Cartwheel has been used as a testbed of the predictions of the collisional
scenario for the formation of ring galaxies \citep{Appleton1996}.  
\citet{Higdon1995} found the current SF, as traced by the \ha\ emission from \hii\ 
regions, is exclusively confined to the ring. The ring is found to be expanding,
with the expansion velocity measured by \citet{Fosbury1977} using optical
longslit spectroscopy being 90~\kms. Subsequent H{\sc i} and Fabry-P\'erot observations
suggest expansion velocity of $\sim$54~\kms\ \citep{Higdon1996, Amram1998}. \citet{Marcum1992} found in the Cartwheel a radial colour gradient that systematically reddens interior to the ring as expected from an ageing population in the wake of the wave. 
\citet{Appleton1997} established such radial colour gradients in other ring galaxies.

Despite the general success of the expanding ring model for the Cartwheel, there 
are still issues that need to be addressed. One of them is regarding a conclusive
evidence of star formation in the past in the zone interior to the ring. 
If the ring was expanding at the currently observed expansion velocities, the
oldest populations would have at the most an age of 450~Myr. \citet{Korchagin2001} 
found that the colour gradients reported by \citet[][]{Appleton1997} 
are too steep for a population age difference expected
between the inner parts and the star-forming ring, and the main cause of the
colour gradient is an underlying old population. \citet{Mayya2005}
carried out 20~cm Very Large Array (VLA) observations to trace
the non-thermal radio continuum (RC) emission from the supernova remnants (SNR) left behind
by the ageing populations in the wake of the wave, but the RC emission
is fainter than expected if the past SFR interior to the ring was similar to the 
presently observed rate of 18~\Msy\ in the ring. In addition, they found the RC 
emission behind the ring is restricted to narrow radial filaments filling only
a small fraction of the azimuthal area. On the other hand, only 10--20\% of the RC 
emission in the ring is thermal, suggesting significant contribution to the RC emission from SNRs.
Due to the relatively large beam size (4~arcsec $=2.4$~kpc) of the 20~cm VLA observations,
it was not able to determine conclusively whether the SNRs are coincident or 
marginally behind (on the inner part of the ring) the position of the present SF as 
traced by \ha.  

\begin{figure*}
\centering\includegraphics[width=2\columnwidth]{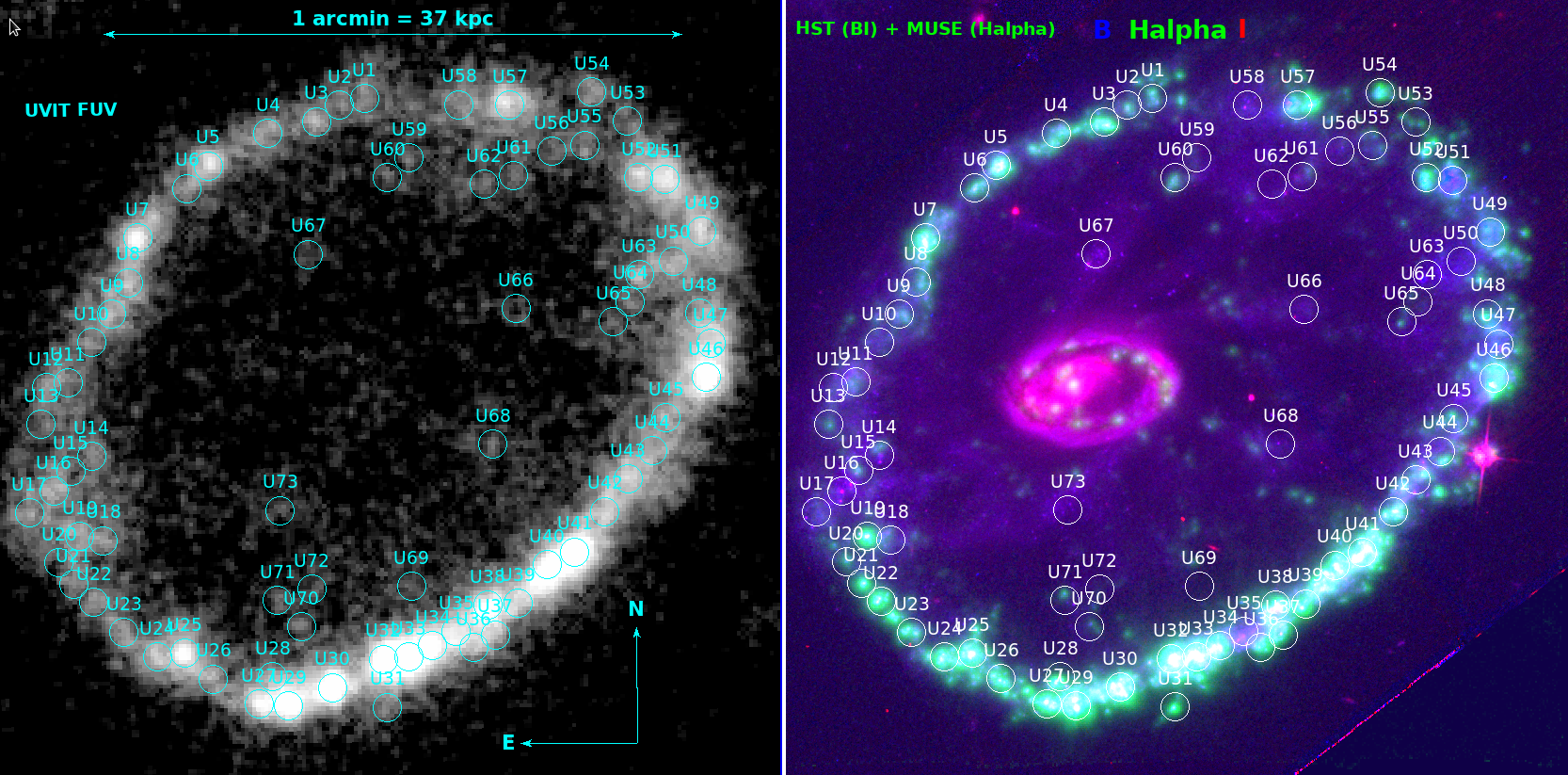}
\caption{(left) FUV compact sources identified on the UVIT FUV image (cyan circles); (right) RGB 
colour-composite image formed using HST/WFPC2 F450W (blue), VLT/MUSE \ha\ (green) and HST/WFPC2 F814W
filters. The circles correspond to 1~arcsec radius, which corresponds to 600~pc at the distance of the Cartwheel.
Note that all the UV sources lying between the inner and outer 
rings belong to one of the spokes of the Cartwheel.
Scale and orientation are indicated.
}
\label{fig:ID}
\end{figure*}

The high angular resolution (0.2~arcsec $=120$~pc) and sensitivity offered by the
Hubble Space Telescope (HST) images offer a direct way to trace the population
of clusters left behind by the advancing wave \citep{Struck1996}. The HST images show more than a 
hundred of stellar clusters, a great majority of them confined to the ring. The
zone interior to the ring has faint star clusters, mostly coincident with the
spokes of the Cartwheel \citep[][]{Hernquist1993}.
The relative faintness of 
these clusters, and the background offered by the spokes and possibly the 
pre-collional disk, has hindered the determination of their ages using optical
colours.

Nevertheless, the lack of clear evidence of past star formation in the wake of 
expanding wave has been a concern to the theories of wave induced star formation.
In order to alleviate this problem, \citet{Renaud2018} carried out a 
detailed hydrodynamical simulations to explain the observed morphology and
star formation in Cartwheel-like galaxies. Using an adaptive mesh refinement code 
they could reach a cell size as small as 6~pc, thus allowing 
them to capture the physical processes coupling the galactic-scale dynamics 
with the star formation activity. They found that the stars formed in the expanding wave are 
carried by the wave as it ploughs through the disc material. 
According to this recent model, the ring hosts all the stars that are formed 
in the expanding wave in the post-collisional disk, and the spokes are the
channels through which some of these stars and collected gas rain back
into the nucleus. Under this scenario, most of the star formation first occurs in the ring before 
this activity is transferred to the spokes and then to the nucleus.

The conclusions from \citet{Renaud2018} models are drastically different 
from that of previous simulations, which predict only recently formed stars 
in the ring, with successively older generation of stars in successively smaller radii.
Tracing the location of stellar populations of up to around a few 100~Myr age
can differentiate between these different predictions.

Images at ultraviolet (UV) wavelengths are ideally suited to trace SF in the recent past without the contaminating effects of light
from any pre-collisional old stars.
This is because, the early and late B stars, which dominate the integrated light
up to around a few hundreds of million years after the death of O stars,
are still hot enough to provide FUV photons. Besides, stars later than 
B-type have strong Balmer jump and contribute insignificantly at the FUV 
wavelengths. The Galex, for the first time, provided far and near UV images
of the Cartwheel. However, its relatively poor spatial resolution inhibited
its use to trace the present and past star forming locations in the Cartwheel.
We have obtained new Far ultraviolet (FUV) images, which are described below, with
point spread function (PSF) of $\sim$1.5~arcsec (900~pc), which is the 
typical angular separation of the star-forming complexes in the Cartwheel.
We here combine these new FUV images with archival images in the optical and infrared to study the locations of present and past star formation in the Cartwheel.

In Section~2, we present the observations used in this work and the catalog of compact FUV sources.
The photometric analysis and extinction determination is presented in Section~3.
In Section~4 we determine the age and mass of the FUV sources using 
population synthesis models in different diagrams.
We discuss the results in Section~5. A summary and our conclusions are given in Section~6.

\section{Observational data and their analysis}

\subsection{New FUV Data}
In this work, we have used data from Ultra-Violet Imaging Telescope (UVIT) onboard ASTROSAT\footnote{https://www.isro.gov.in/AstroSat.html}, which is the first Indian Space Observatory launched by the Indian Space Research Organisation(ISRO). UVIT consists of two identical 38 cm co-aligned telescopes, one for the FUV channel from 130-180~nm and the other for the near-ultraviolet (NUV) from 200-300~nm and VIS channel from 320-550~nm, with a ﬁeld of view of 28~arcmin at a spatial  resolution $<$ 1.5~arcsec. The FUV and the NUV detectors are operated in the high gain photon counting mode. In contrast, the VIS detector is operated in the low gain integration mode. The FUV and NUV channels have ﬁlters and gratings, whereas the VIS channel has ﬁlters. The description of UVIT telescope and calibration details can be found in \citet{Kumar2012, Tandon2017, Tandon2020}.

The Cartwheel galaxy was observed during two epochs: 11 November 2018 and 15 December 2018 (see Table~\ref{tab:obslog} for details of observations) in FUV data in the F148W (CaF2; mean $\lambda$ 1481 \AA) filter. The orbit-wise dataset was processed using the Level-2 (L2) pipeline which gives data products that are readily usable for science, including intensity and error on intensity maps and effective exposure maps. The L2 processing pipeline has been used for the data reduction, which includes: 1) extraction of drift in the pointing of the spacecraft and disturbances in pointing due to internal movements and 2) application of various corrections to measured position in the detector for each photon such as differential pointing with respect to a reference frame for shift and adds operation, systematic effects and artefacts in the optics of the telescopes and detectors, exposure tracking on the sky, alignment of sky products from multi-episode exposures to generate a consolidated set and astrometry \citep{Ghosh2022, Ghosh2021}. The pipeline astrometry is sometimes inaccurate, so we applied our astrometric correction to orbit-wise images using the GALEX FUV/NUV tiles. The orbit-wise images are median combined to produce the final deep image shown in the left panel of Figure~\ref{fig:ID}.

\begin{table}
\caption{Log of FUV imaging observations in the Cartwheel.}\label{tab:obslog} 
\begin{tabular}{cccc}
\hline
ID   & Date       & UT start    & Exp. time \\
     &(yyyy/mm/dd)& (hh:mm:ss) & (s) \\
\hline
a1 &  2018/11/11  &   05:10:51  &   614 \\
a2 &  2018/11/11  &   05:22:14  &   900 \\
a3 &  2018/11/11  &   08:25:44  &   1211 \\
a4 &  2018/11/11  &   10:03:11  &   1583 \\
a5 &  2018/11/11  &   11:40:38  &   1740 \\
a6 &  2018/11/11  &   13:18:04  &   602 \\
b1 &  2018/12/15  &   14:08:53  &   1287 \\
b2 &  2018/12/15  &   15:46:19  &   1287 \\
b3 &  2018/12/15  &   17:23:45  &   1288 \\
b4 &  2018/12/15  &   19:01:12  &   1285 \\
b5 &  2018/12/15  &   20:43:42  &   1001 \\
b6 &  2018/12/15  &   22:28:12  &   506 \\
\hline
\end{tabular}
\end{table}

\subsection{Compact sources detected on the FUV image}

The most easily identifiable characteristic in the UVIT FUV image is the star-forming ring, which is resolved into several individual knots of UV emission. Additionally FUV emission is detected from a few sources along the spokes that connects the outer ring and the nucleus. The nucleus and the inner ring are completely absent in the UV image. 

We identified 73 sources on the FUV image, with $\sim$60 of them belonging to the outer ring or close to it. These sources are identified in Figure~\ref{fig:ID} and their equatorial coordinates are given in columns 2 and 3 of Table~\ref{tab:fuvmag}.

\begin{table}
\small\addtolength{\tabcolsep}{-1.5pt}
\scriptsize
\caption{Catalog of compact FUV sources in the Cartwheel$^\dagger$}
\label{tab:fuvmag} 
\begin{tabular}{rcccccc}
\hline
     & \multicolumn{2}{c}{Coordinates (J2000)} & \multicolumn{4}{c}{Photometry} \\
  ID & R.A.       & Dec.        &$m_{\rm FUV}$& e$m_{\rm FUV}$ & $M_{\rm FUV}^0$ & e$M_{\rm FUV}$ \\
     & (h)        & (deg)       & (mag)    & (mag)   & (mag)     &  (mag) \\
\hline
1 &    9.418629 & $-$33.70820 &   19.897 &   0.063 & $-$17.109 &   0.118 \\
2 &    9.419499 & $-$33.70839 &   19.959 &   0.065 & $-$16.776 &   0.119 \\
3 &    9.420283 & $-$33.70887 &   19.612 &   0.054 & $-$17.372 &   0.060 \\
4 &    9.421943 & $-$33.70920 &   19.599 &   0.054 & $-$17.525 &   0.071 \\
5 &    9.424007 & $-$33.71012 &   18.855 &   0.037 & $-$18.149 &   0.042 \\
6 &    9.424725 & $-$33.71077 &   19.640 &   0.055 & $-$17.337 &   0.091 \\
7 &    9.426428 & $-$33.71219 &   18.557 &   0.032 & $-$17.819 &   0.035 \\
8 &    9.426749 & $-$33.71346 &   19.175 &   0.043 & $-$17.594 &   0.091 \\
9 &    9.427325 & $-$33.71437 &   19.311 &   0.046 & $-$17.305 &   0.161 \\
10 &    9.428013 & $-$33.71517 &   19.885 &   0.063 & $-$16.731 &   0.667 \\
11 &    9.428822 & $-$33.71632 &   19.840 &   0.061 & $-$16.776 &   0.378 \\
12 &    9.429570 & $-$33.71647 &   19.746 &   0.058 & $-$16.870 &   0.735 \\
13 &    9.429746 & $-$33.71752 &   20.291 &   0.078 & $-$16.325 &   0.433 \\
14 &    9.428008 & $-$33.71842 &   20.033 &   0.068 & $-$16.583 &   0.468 \\
15 &    9.428725 & $-$33.71885 &   19.894 &   0.063 & $-$16.722 &   1.256 \\
16 &    9.429308 & $-$33.71943 &   19.369 &   0.048 & $-$17.248 &   0.048 \\
17 &    9.430153 & $-$33.72004 &   19.674 &   0.056 & $-$16.942 &   0.905 \\
18 &    9.427604 & $-$33.72084 &   19.742 &   0.058 & $-$16.984 &   0.115 \\
19 &    9.428412 & $-$33.72073 &   19.383 &   0.048 & $-$16.826 &   0.052 \\
20 &    9.429137 & $-$33.72147 &   20.064 &   0.069 & $-$16.588 &   0.120 \\
21 &    9.428625 & $-$33.72208 &   19.933 &   0.064 & $-$16.435 &   0.072 \\
22 &    9.427942 & $-$33.72260 &   19.808 &   0.060 & $-$17.062 &   0.063 \\
23 &    9.426895 & $-$33.72347 &   19.653 &   0.055 & $-$17.053 &   0.069 \\
24 &    9.425757 & $-$33.72417 &   18.924 &   0.038 & $-$17.556 &   0.040 \\
25 &    9.424816 & $-$33.72406 &   18.427 &   0.030 & $-$18.083 &   0.031 \\
26 &    9.423825 & $-$33.72480 &   19.465 &   0.050 & $-$17.500 &   0.055 \\
27 &    9.422257 & $-$33.72552 &   18.373 &   0.030 & $-$18.081 &   0.031 \\
28 &    9.421796 & $-$33.72474 &   18.830 &   0.037 & $-$17.459 &   0.040 \\
29 &    9.421250 & $-$33.72557 &   17.851 &   0.023 & $-$18.623 &   0.024 \\
30 &    9.419742 & $-$33.72504 &   17.870 &   0.023 & $-$18.669 &   0.024 \\
31 &    9.417846 & $-$33.72561 &   20.197 &   0.086 & $-$16.419 &   0.097 \\
32 &    9.417995 & $-$33.72424 &   17.432 &   0.019 & $-$19.224 &   0.019 \\
33 &    9.417125 & $-$33.72417 &   17.338 &   0.018 & $-$19.149 &   0.018 \\
34 &    9.416291 & $-$33.72385 &   17.846 &   0.023 & $-$18.712 &   0.025 \\
35 &    9.415487 & $-$33.72344 &   18.282 &   0.028 & $-$18.334 &   0.061 \\
36 &    9.414900 & $-$33.72390 &   18.781 &   0.036 & $-$17.906 &   0.046 \\
37 &    9.414149 & $-$33.72354 &   18.802 &   0.036 & $-$17.714 &   0.044 \\
38 &    9.414399 & $-$33.72269 &   18.126 &   0.026 & $-$18.320 &   0.029 \\
39 &    9.413379 & $-$33.72264 &   18.603 &   0.033 & $-$17.899 &   0.034 \\
40 &    9.412367 & $-$33.72154 &   17.819 &   0.022 & $-$18.469 &   0.023 \\
41 &    9.411412 & $-$33.72117 &   17.597 &   0.020 & $-$18.886 &   0.021 \\
42 &    9.410375 & $-$33.72002 &   18.834 &   0.037 & $-$18.154 &   0.045 \\
43 &    9.409595 & $-$33.71909 &   18.795 &   0.036 & $-$17.821 &   0.066 \\
44 &    9.408746 & $-$33.71827 &   19.177 &   0.043 & $-$17.439 &   0.118 \\
45 &    9.408283 & $-$33.71734 &   19.247 &   0.045 & $-$17.369 &   0.119 \\
46 &    9.406903 & $-$33.71618 &   17.955 &   0.024 & $-$18.782 &   0.028 \\
47 &    9.406745 & $-$33.71520 &   18.392 &   0.030 & $-$18.224 &   0.091 \\
48 &    9.407125 & $-$33.71434 &   19.120 &   0.042 & $-$18.420 &   0.081 \\
49 &    9.407049 & $-$33.71201 &   18.784 &   0.036 & $-$17.832 &   0.087 \\
50 &    9.408042 & $-$33.71285 &   19.242 &   0.045 & $-$17.374 &   0.045 \\
51 &    9.408321 & $-$33.71054 &   18.477 &   0.031 & $-$18.139 &   0.138 \\
52 &    9.409221 & $-$33.71047 &   18.804 &   0.036 & $-$17.941 &   0.054 \\
53 &    9.409595 & $-$33.70886 &   20.150 &   0.072 & $-$16.466 &   0.173 \\
54 &    9.410829 & $-$33.70803 &   19.848 &   0.061 & $-$16.431 &   0.088 \\
55 &    9.411078 & $-$33.70954 &   20.110 &   0.071 & $-$16.506 &   0.374 \\
56 &    9.412204 & $-$33.70972 &   20.511 &   0.089 & $-$16.105 &   0.089 \\
57 &    9.413642 & $-$33.70840 &   18.802 &   0.036 & $-$17.846 &   0.044 \\
58 &    9.415383 & $-$33.70839 &   19.732 &   0.058 & $-$16.884 &   0.058 \\
59 &    9.417100 & $-$33.70991 &   20.929 &   0.114 & $-$15.688 &   0.913 \\
60 &    9.417846 & $-$33.71047 &   21.219 &   0.136 & $-$17.043 &   0.276 \\
61 &    9.413507 & $-$33.71043 &   20.864 &   0.120 & $-$15.752 &   0.120 \\
62 &    9.414536 & $-$33.71066 &   20.827 &   0.107 & $-$15.789 &   0.107 \\
63 &    9.409200 & $-$33.71323 &   19.560 &   0.053 & $-$17.056 &   0.053 \\
64 &    9.409507 & $-$33.71402 &   20.362 &   0.081 & $-$16.254 &   0.081 \\
65 &    9.410083 & $-$33.71459 &   20.865 &   0.109 & $-$15.751 &   0.109 \\
66 &    9.413428 & $-$33.71422 &   21.761 &   0.196 & $-$14.855 &   0.196 \\
67 &    9.420567 & $-$33.71265 &   21.625 &   0.179 & $-$14.991 &   0.179 \\
68 &    9.414232 & $-$33.71809 &   21.140 &   0.130 & $-$15.476 &   0.130 \\
69 &    9.417021 & $-$33.72215 &   21.156 &   0.134 & $-$15.460 &   0.757 \\
70 &    9.420808 & $-$33.72330 &   20.880 &   0.110 & $-$15.736 &   0.767 \\
71 &    9.421629 & $-$33.72254 &   21.372 &   0.151 & $-$15.244 &   0.477 \\
72 &    9.420445 & $-$33.72223 &   21.321 &   0.146 & $-$15.295 &   0.146 \\
73 &    9.421537 & $-$33.71997 &   21.446 &   0.158 & $-$15.170 &   0.158 \\
\hline
\end{tabular}
$^\dagger$Vega magnitudes in 1.5~arcsec radius apertures. $M_{\rm FUV^0}$ is the absolute magnitude corrected for extinction using \citet{Cardelli1989} extinction curve.
\end{table}

\begin{table}
\small\addtolength{\tabcolsep}{-4pt}
\caption{Description of the multiwavelenght data used in this work.}
\label{tab:multiband}
\begin{tabular}{cccccccc}
\hline
Mission &  Filter & $\lambda_{\rm c}$ & $\Delta\lambda$& FWHM & T$_{\rm exp}$ & ZP & $m_{\rm lim}$\\
                         &         & (nm)   & (nm)   & (arcsec) & (s) & Vega    & 3$\sigma$    \\
                     (1) &  (2)    & (3)     & (4)     & (5)   & (6)  & (7)   & (8)    \\
\hline
Astrosat/{\sc uvit}      & FUV     & 148.5 & 40.2  & 1.5   & 13304   & 16.00 & 22.79 \\ 
HST/{\sc wfpc2}          &  F450W  & 454.5 & 78.9  & 0.20  & 3200 & 24.30 & 24.59 \\ 
HST/{\sc wfpc2}          &  F814W  & 793.0 & 164 & 0.20  & 3200 & 23.65 & 22.95 \\ 
VLT/{\sc muse}$^\dagger$ & \ha\    & 676.4    & 2       & 0.60  & 4030 & ---   & 1.4   \\ 
VLT/{\sc muse}$^\dagger$ & \hb\    & 500.8    & 2       & 0.60  & 4030 & ---   & 2.5   \\ 
VLT/{\sc isaac}          & K       & 2200   & 350      & 0.60  & 4$\times$31 & 24.19 & 21.22 \\ 
\hline
\end{tabular}
$^\dagger$ The MUSE datacube is used to generate maps in nebular lines \hb\ and \ha\ at their observed wavelengths ($\lambda_{\rm c}$) integrating the line flux in a window of $\Delta\lambda$, and subtracting the average continuum measured at 37.5~\AA\ on either side of the $\lambda_{\rm c}$.
The last column for these filters contains 3-$\sigma$ limiting fluxes in units of $\times 10^{-17}$ \ecms.
\end{table}

\subsection{Multi-band archival data}

We used the available archival data to complement the FUV image from UVIT. These include the WFPC2 images in the F450W and F814W bands from the HST Legacy Archive, MUSE spectral cube and the K-band image from the ESO archive. A summary of the data used in this work is given in Table~\ref{tab:multiband}. The tabulated values include the central wavelength and bandwidth, the Full Width at Half Maximum (FWHM) of the Point Spread Function (PSF), exposure time, the zeropoints and the 3-$\sigma$ limiting magnitudes of each filter. 
All the magnitudes in this work are presented in the Vega system and accordingly the tabulated zeropoints are also in the Vega system. We used the FUV AB zeropoint pf 18.08 from \citet[][]{Tandon2020}, which was converted to Vega system using the relation $m_{\rm Vega} = m_{\rm AB} - m_{\rm AB}$(Vega), where $m_{\rm AB}$(Vega)=2.085 is the AB magnitude of Vega in the FUV filter, which was taken from \citet{Willmer2018} for the GALEX FUV filter.  

\subsubsection{Optical continuum and Near infrared images}

The Cartwheel was observed by HST in its WFPC2 camera in F450W and F814W bands which trace the continuum emission from the stars. These images have a Point Spread Function (PSF) of 0.2~arcsec, which corresponds to 120~pc of physical resolution. The star-forming outer ring is resolved into multiple stellar clusters at this resolution. The zeropoints for the HST filters were taken from the image headers, which are found to agree within 0.20~mag with the values we obtained from synthetic photometry on the MUSE spectra of bright unsaturated sources common in the HST and MUSE field of views.
In an earlier work \citep[][]{Barway2020}, we have analysed the K-band images in the ESO archive to study the morphology of the central bar in the Cartwheel, which are used in the present work to carry out photometry on UV selected regions. 

\subsubsection{\ha\ and \hb\ images}
The MUSE datacube was used to generate \ha\ and \hb\ emission line maps. The emission line maps are generated by integrating in filters of 20~\AA\ width around the redshifted wavelengths of \ha\ and \hb\ and subtracting an average of the continuum in windows of 20~\AA\ width. The continuum windows are centered at 37.5~\AA\ on either side of the line, and are free of any emission lines. 
The archival MUSE datacube is already flux calibrated, and does not require to use zeropoints, which is the reason for the blank values in column~7 of Table~\ref{tab:multiband} for the \ha\ and \hb\ filters. The limiting magnitudes are calculated in apertures of 1.5~arcsec radius.

\section{Photometric analysis}

\subsection{Aperture photometry of compact sources}

Though the FUV image provided by the UVIT instrument is the best in spatial resolution at UV wavelengths, it still corresponds to a physical scale of 900~pc given the relatively large distance to the Cartwheel. This scale, nevertheless is just sufficient to separate the star forming complexes in the ring of the Cartwheel. We hence chose apertures of 1.5~arcsec radius to carry out photometry in all bands. In order to take into account the vastly different FWHM of the point spread functions of the archival images, we resampled all the archival images to the pixel scale of the UVIT image (0.417 arcsec/pixel) and smoothed them to the resolution of the UVIT image. All the photometry was carried out on these resampled images.

We used the {\sc iraf} task {\sc phot} to carry out aperture photometry in concentric circular apertures of radius between 0.5 to 2.5 arcsec in steps of 0.5~arcsec. 
We used the median value of an annular zone of inner radius and width of 10 and 5~arcsec, respectively, centered on the object as the local background value, and subtracted it from each pixel value, to obtain the magnitude of the source. Note that the chosen annulus is large enough so that the median value is not affected from any contribution from the neighbouring sources in the ring.
The instrumental magnitudes are converted to Vega system of magnitudes using the zeropoints in column~7 of Table~\ref{tab:multiband}. The photometric errors include the photon noise from the source and the background. The magnitude as a function of aperture radius grows even at the largest aperture of 2.5~arcsec we have used. This is expected for two reasons, (1) the PSF of the smoothed images used for photometry extends beyond the aperture of 2.5~arcsec aperture, and (2) a neighbouring region starts contributing to large apertures due to the crowding of regions. It can be inferred from Figure~\ref{fig:ID} that an aperture of 1.5~arcsec radius encloses a complex without getting contribution from neighbouring in almost all cases. We hence chose radius of 1.5~arcsec as the optimum aperture in this study. We found that the colour remains steady as a function of aperture size, in spite of the growth of the magnitude. As we will discuss shortly, the age of the population, which we aim to derive using our dataset, is sensitive to colour rather than magnitude and our results are only weakly dependent on the choice of aperture size.

In columns 4--7 of Table~\ref{tab:fuvmag}, we give FUV magnitudes in the Vega system for the identified UV sources. The $m_{\rm FUV}$ is the observed apparent magnitude, whereas the $M_{\rm UV}^0$ is the absolute magnitude corrected for interstellar extinction discussed below. The error on $M_{\rm UV}$ includes error on extinction value at the FUV wavelength as well as the photometric errors.
Table~\ref{tab:obsdata} gives the multi-band photometric properties of the identified UV sources. 

\begin{table*}
\addtolength{\tabcolsep}{-1pt}
\scriptsize
\caption{Multi-band photometric data of UV sources in the Cartwheel.}\label{tab:obsdata} 
\begin{tabular}{rrrrrrrrrrrrrrrrrrr}
\hline
 ID &   B &  eB &  U$-$B & eU$-$B & B$-$I & eBI & B$-$K & eBK &  F(\ha)     & \multicolumn{2}{c}{SNR} & \ha/\hb\      & $A_V$ & e$A_V$ & \multicolumn{2}{c}{\ha\ [\AA]} & \multicolumn{2}{c}{\hb\ [\AA]} \\
 ID &     &     &        &        &       &     &       &     &            & \ha\     &         \hb\ &            &       &  & EW & eEW & EW & eEW \\
  (1) & (2) & (3) & (4) & (5) & (6) & (7) & (8) & (9) & (10) & (11) & (12) & (13) & (14) & (15) & (16) & (17) & (18) & (19) \\
\hline
   1 &   21.12 &   0.04 & $-$1.22 &   0.08 &    0.73 &   0.10 &    1.76 &   0.17 &  1.028e-15 &   221 &    31 &    3.42 &    0.54 &   0.10 &      83 &      5 &      12 &      1\\
   2 &   21.06 &   0.04 & $-$1.10 &   0.08 &    0.98 &   0.08 &    2.07 &   0.12 &  9.998e-16 &   215 &    31 &    3.31 &    0.44 &   0.10 &      75 &      4 &      12 &      1\\
   3 &   20.43 &   0.02 & $-$0.82 &   0.06 &    0.53 &   0.06 &    1.80 &   0.09 &  3.764e-15 &   809 &   126 &    3.41 &    0.54 &   0.03 &     255 &     14 &      40 &      2\\
   4 &   20.94 &   0.05 & $-$1.34 &   0.07 &    0.69 &   0.10 &    1.95 &   0.12 &  2.083e-15 &   447 &    67 &    3.47 &    0.59 &   0.05 &     199 &     16 &      28 &      2\\
   5 &   20.01 &   0.02 & $-$1.16 &   0.04 &    0.66 &   0.04 &    1.84 &   0.05 &  4.836e-15 &  1040 &   159 &    3.42 &    0.54 &   0.02 &     198 &      6 &      32 &      1\\
   6 &   20.89 &   0.04 & $-$1.25 &   0.07 &    0.83 &   0.08 &    2.03 &   0.09 &  1.369e-15 &   295 &    43 &    3.41 &    0.53 &   0.07 &     110 &      7 &      16 &      1\\
   7 &   20.08 &   0.02 & $-$1.53 &   0.04 &    0.54 &   0.05 &    1.82 &   0.05 &  5.719e-15 &  1231 &   206 &    3.17 &    0.31 &   0.02 &     284 &     11 &      46 &      1\\
   8 &   20.93 &   0.05 & $-$1.76 &   0.06 &    0.62 &   0.11 &    1.98 &   0.12 &  1.205e-15 &   259 &    39 &    3.33 &    0.46 &   0.08 &     124 &     12 &      16 &      1\\
   9 &   20.94 &   0.05 & $-$1.63 &   0.07 &    0.64 &   0.11 &    1.47 &   0.22 &  7.924e-16 &   171 &    20 &    3.82 &    0.40 &   0.16 &      75 &      7 &       8 &      1\\
  10 &   21.23 &   0.06 & $-$1.34 &   0.09 &    0.82 &   0.12 &    2.36 &   0.15 &  3.394e-16 &    73 &     4 &    4.78 &    0.40 &   0.75 &      35 &      3 &       2 &      0\\
  11 &   20.72 &   0.04 & $-$0.87 &   0.07 &    0.84 &   0.08 &    1.40 &   0.22 &  5.474e-16 &   118 &     8 &    4.38 &    0.40 &   0.40 &      35 &      2 &       2 &      0\\
  12 &   20.96 &   0.05 & $-$1.22 &   0.08 &    0.61 &   0.13 &    2.01 &   0.15 &  3.774e-16 &    81 &     4 &    4.97 &    0.40 &   0.85 &      36 &      4 &       1 &      0\\
  13 &   21.68 &   0.11 & $-$1.39 &   0.13 &    0.46 &   0.27 &    1.59 &   0.32 &  3.806e-16 &    82 &     7 &    4.90 &    0.40 &   0.46 &      78 &     16 &       6 &      1\\
  14 &   21.21 &   0.05 & $-$1.18 &   0.09 &    0.90 &   0.10 &    1.89 &   0.19 &  2.896e-16 &    62 &     6 &    3.21 &    0.40 &   0.40 &      29 &      2 &       3 &      0\\
  15 &   20.82 &   0.04 & $-$0.93 &   0.07 &    0.97 &   0.07 &    2.02 &   0.11 &  4.125e-16 &    89 &     2 &    5.79 &    0.40 &   1.70 &      27 &      1 &       1 &      0\\
  16 &   20.53 &   0.03 & $-$1.16 &   0.06 &    0.93 &   0.06 &    2.08 &   0.08 &  1.801e-16 &    39 &     3 &    1.82 &    0.40 &   0.00 &       8 &      0 &       1 &      0\\
  17 &   20.99 &   0.05 & $-$1.31 &   0.08 &    0.76 &   0.10 &    1.62 &   0.19 &  2.899e-16 &    62 &     3 &    4.30 &    0.40 &   1.08 &      25 &      2 &       1 &      0\\
  18 &   20.99 &   0.05 & $-$1.25 &   0.08 &    0.82 &   0.10 &    2.12 &   0.10 &  9.955e-16 &   214 &    31 &    3.31 &    0.44 &   0.10 &      86 &      7 &      13 &      1\\
  19 &   20.41 &   0.03 & $-$1.03 &   0.06 &    0.43 &   0.07 &    1.53 &   0.11 &  4.476e-15 &   963 &   166 &    3.11 &    0.25 &   0.02 &     303 &     16 &      55 &      3\\
  20 &   21.32 &   0.06 & $-$1.25 &   0.09 &    0.73 &   0.12 &    1.82 &   0.19 &  9.679e-16 &   208 &    31 &    3.28 &    0.41 &   0.10 &      99 &      7 &      16 &      1\\
  21 &   20.96 &   0.03 & $-$1.03 &   0.07 &    0.60 &   0.08 &    1.51 &   0.17 &  2.748e-15 &   591 &    99 &    3.17 &    0.31 &   0.03 &     240 &     12 &      44 &      2\\
  22 &   20.67 &   0.01 & $-$0.86 &   0.06 &    0.64 &   0.03 &    1.44 &   0.11 &  4.785e-15 &  1029 &   165 &    3.37 &    0.49 &   0.02 &     364 &      9 &      64 &      1\\
  23 &   20.81 &   0.02 & $-$1.15 &   0.06 &    0.81 &   0.03 &    1.38 &   0.13 &  2.222e-15 &   477 &    75 &    3.30 &    0.43 &   0.04 &     162 &      4 &      28 &      1\\
  24 &   20.11 &   0.01 & $-$1.18 &   0.04 &    0.59 &   0.03 &    1.45 &   0.06 &  6.887e-15 &  1477 &   247 &    3.21 &    0.35 &   0.01 &     314 &      7 &      56 &      1\\
  25 &   19.70 &   0.01 & $-$1.27 &   0.03 &    0.46 &   0.03 &    1.53 &   0.04 &  1.078e-14 &  2308 &   386 &    3.22 &    0.36 &   0.01 &     373 &      9 &      63 &      1\\
  26 &   20.74 &   0.03 & $-$1.27 &   0.06 &    0.62 &   0.06 &    1.84 &   0.07 &  4.143e-15 &   887 &   139 &    3.40 &    0.53 &   0.02 &     318 &     13 &      51 &      2\\
  27 &   19.61 &   0.01 & $-$1.24 &   0.03 &    0.20 &   0.02 &    1.46 &   0.02 &  1.296e-14 &  2769 &   470 &    3.20 &    0.34 &   0.01 &     571 &     12 &     102 &      2\\
  28 &   20.69 &   0.02 & $-$1.86 &   0.04 &    0.61 &   0.05 &    1.88 &   0.05 &  5.150e-15 &  1104 &   189 &    3.14 &    0.28 &   0.02 &     387 &     15 &      69 &      2\\
  29 &   19.35 &   0.01 & $-$1.50 &   0.02 &    0.36 &   0.02 &    1.74 &   0.02 &  1.728e-14 &  3690 &   621 &    3.21 &    0.35 &   0.01 &     468 &      5 &      83 &      1\\
  30 &   19.39 &   0.01 & $-$1.52 &   0.02 &    0.59 &   0.01 &    1.90 &   0.01 &  1.394e-14 &  2986 &   494 &    3.23 &    0.37 &   0.01 &     327 &      4 &      57 &      1\\
  31 &   22.84 &   0.18 & $-$2.65 &   0.20 &    1.20 &   0.28 &    3.34 &   0.21 &  1.797e-15 &   384 &    70 &    3.03 &    0.17 &   0.05 &     614 &     102 &     0 &      9\\
  32 &   18.44 &   0.00 & $-$1.01 &   0.02 &    0.18 &   0.01 &    1.46 &   0.01 &  5.141e-14 & 11046 &  1839 &    3.28 &    0.41 &   0.00 &     761 &      6 &     136 &      1\\
  33 &   18.86 &   0.01 & $-$1.52 &   0.02 &    0.59 &   0.01 &    1.90 &   0.01 &  1.921e-14 &  4129 &   684 &    3.21 &    0.35 &   0.00 &     282 &      2 &      49 &      0\\
  34 &   19.46 &   0.01 & $-$1.61 &   0.02 &    0.62 &   0.02 &    1.88 &   0.02 &  9.209e-15 &  1980 &   322 &    3.24 &    0.38 &   0.01 &     221 &      3 &      38 &      0\\
  35 &   19.98 &   0.01 & $-$1.69 &   0.03 &    0.90 &   0.02 &    2.11 &   0.03 &  2.032e-15 &   437 &    57 &    3.43 &    0.40 &   0.06 &      60 &      1 &       8 &      0\\
  36 &   20.49 &   0.02 & $-$1.71 &   0.04 &    0.71 &   0.05 &    1.81 &   0.06 &  3.236e-15 &   695 &   108 &    3.29 &    0.43 &   0.03 &     154 &      4 &      25 &      1\\
  37 &   20.49 &   0.03 & $-$1.69 &   0.04 &    0.74 &   0.05 &    2.01 &   0.05 &  3.604e-15 &   774 &   125 &    3.23 &    0.36 &   0.03 &     168 &      4 &      30 &      1\\
  38 &   19.64 &   0.01 & $-$1.51 &   0.03 &    0.78 &   0.02 &    2.23 &   0.02 &  6.489e-15 &  1395 &   226 &    3.20 &    0.34 &   0.01 &     165 &      2 &      29 &      0\\
  39 &   20.04 &   0.02 & $-$1.44 &   0.04 &    0.52 &   0.04 &    1.87 &   0.04 &  9.121e-15 &  1961 &   328 &    3.22 &    0.36 &   0.01 &     371 &      8 &      68 &      1\\
  40 &   19.31 &   0.01 & $-$1.49 &   0.02 &    0.54 &   0.02 &    1.90 &   0.02 &  1.477e-14 &  3175 &   543 &    3.14 &    0.28 &   0.01 &     337 &      5 &      60 &      1\\
  41 &   19.13 &   0.01 & $-$1.53 &   0.02 &    0.68 &   0.02 &    2.07 &   0.01 &  1.539e-14 &  3310 &   547 &    3.21 &    0.35 &   0.01 &     248 &      2 &      45 &      0\\
  42 &   20.10 &   0.01 & $-$1.27 &   0.04 &    0.64 &   0.03 &    2.09 &   0.03 &  3.721e-15 &   800 &   120 &    3.41 &    0.54 &   0.03 &     149 &      3 &      23 &      0\\
  43 &   20.10 &   0.01 & $-$1.31 &   0.04 &    0.80 &   0.03 &    2.10 &   0.04 &  1.972e-15 &   424 &    56 &    3.53 &    0.40 &   0.06 &      72 &      2 &      10 &      0\\
  44 &   20.40 &   0.02 & $-$1.23 &   0.05 &    0.93 &   0.03 &    2.26 &   0.06 &  1.193e-15 &   256 &    28 &    3.69 &    0.40 &   0.11 &      38 &      1 &       5 &      0\\
  45 &   20.54 &   0.02 & $-$1.29 &   0.05 &    0.85 &   0.04 &    1.80 &   0.11 &  1.134e-15 &   244 &    28 &    3.73 &    0.40 &   0.11 &      50 &      1 &       7 &      0\\
  46 &   19.50 &   0.01 & $-$1.55 &   0.03 &    0.34 &   0.02 &    1.64 &   0.06 &  6.298e-15 &  1352 &   207 &    3.31 &    0.44 &   0.02 &     132 &      2 &      22 &      0\\
  47 &   20.13 &   0.02 & $-$1.74 &   0.04 &    0.47 &   0.04 &    1.60 &   0.12 &  1.480e-15 &   318 &    36 &    3.67 &    0.40 &   0.09 &      51 &      1 &       6 &      0\\
  48 &   20.57 &   0.03 & $-$1.45 &   0.05 &    0.41 &   0.05 &    1.81 &   0.14 &  1.597e-15 &   343 &    45 &    3.65 &    0.74 &   0.07 &      86 &      3 &      12 &      0\\
  49 &   20.20 &   0.02 & $-$1.42 &   0.04 &    0.31 &   0.04 &    1.82 &   0.08 &  1.512e-15 &   325 &    39 &    3.60 &    0.40 &   0.08 &      57 &      1 &       7 &      0\\
  50 &   20.61 &   0.03 & $-$1.36 &   0.05 &    0.56 &   0.04 &    2.16 &   0.08 &  2.517e-16 &    54 &     3 &    2.73 &    0.40 &   0.00 &      13 &      1 &       1 &      0\\
  51 &   20.03 &   0.01 & $-$1.55 &   0.03 &    0.14 &   0.03 &    1.87 &   0.05 &  1.105e-15 &   237 &    23 &    3.68 &    0.40 &   0.14 &      34 &      1 &       3 &      0\\
  52 &   20.38 &   0.02 & $-$1.58 &   0.04 &   -0.01 &   0.05 &    1.74 &   0.07 &  2.348e-15 &   504 &    76 &    3.32 &    0.45 &   0.04 &     114 &      3 &      18 &      0\\
  53 &   21.98 &   0.03 & $-$1.83 &   0.08 &    0.60 &   0.10 &    2.21 &   0.20 &  5.566e-16 &   120 &    19 &    3.08 &    0.40 &   0.16 &      93 &      3 &      18 &      1\\
  54 &   21.44 &   0.02 & $-$1.59 &   0.06 &    0.59 &   0.06 &    1.69 &   0.14 &  1.376e-15 &   296 &    49 &    3.14 &    0.28 &   0.06 &     177 &      4 &      32 &      1\\
  55 &   21.29 &   0.03 & $-$1.18 &   0.08 &    0.67 &   0.06 &    1.67 &   0.15 &  3.583e-16 &    77 &     8 &    3.44 &    0.40 &   0.39 &      30 &      1 &       4 &      0\\
  56 &   21.49 &   0.06 & $-$0.98 &   0.11 &    0.97 &   0.08 &    2.07 &   0.13 &  1.373e-16 &    30 &     3 &    2.61 &    0.40 &   0.00 &      18 &      2 &       2 &      1\\
  57 &   20.09 &   0.02 & $-$1.29 &   0.04 &    0.15 &   0.05 &    1.87 &   0.04 &  3.585e-15 &   770 &   121 &    3.28 &    0.41 &   0.03 &     160 &      5 &      26 &      1\\
  58 &   20.92 &   0.04 & $-$1.19 &   0.07 &    0.75 &   0.08 &    1.85 &   0.11 &  1.404e-16 &    30 &     3 &    1.99 &    0.40 &   0.00 &      11 &      1 &       1 &      0\\
  59 &   21.96 &   0.09 & $-$1.03 &   0.15 &    0.77 &   0.20 &    2.96 &   0.13 &  1.196e-16 &    26 &     3 &    2.88 &    0.40 &   0.46 &      25 &      4 &       3 &      1\\
  60 &   22.22 &   0.13 & $-$1.00 &   0.19 &    0.40 &   0.36 &    2.52 &   0.21 &  4.600e-16 &    99 &    13 &    3.97 &    1.01 &   0.25 &     162 &     43 &      20 &      4\\
  61 &   21.75 &   0.07 & $-$0.89 &   0.14 &    1.16 &   0.09 &    2.43 &   0.12 &  1.007e-16 &    22 &     3 &    2.12 &    0.40 &   0.00 &      15 &      1 &       2 &      1\\
  62 &   21.98 &   0.08 & $-$1.15 &   0.13 &    1.15 &   0.10 &    2.39 &   0.14 &  7.310e-18 &     2 &     3 &    0.17 &    0.40 &   0.00 &       1 &      1 &       3 &      1\\
  63 &   20.64 &   0.03 & $-$1.08 &   0.06 &    0.31 &   0.05 &    2.38 &   0.07 &  5.414e-17 &    12 &     3 &    0.49 &    0.40 &   0.00 &       2 &      0 &       1 &      0\\
  64 &   21.98 &   0.09 & $-$1.62 &   0.12 &    0.62 &   0.16 &    3.04 &   0.17 &  1.343e-16 &    29 &     6 &    1.87 &    0.40 &   0.00 &      23 &      3 &       5 &      1\\
  65 &   22.18 &   0.10 & $-$1.32 &   0.15 &    0.73 &   0.16 &    2.51 &   0.28 &  1.545e-16 &    33 &     8 &    1.92 &    0.40 &   0.00 &      44 &      7 &      10 &      2\\
  66 &   22.32 &   0.08 & $-$0.56 &   0.21 &    1.16 &   0.13 &    2.91 &   0.19 &  2.428e-17 &     5 &     3 &    0.65 &    0.40 &   0.00 &       7 &      2 &       4 &      1\\
  67 &   21.50 &   0.05 & $-$-0.13 &  0.19 &    1.28 &   0.07 &    2.74 &   0.07 &  5.675e-18 &     1 &     3 &    0.10 &    0.40 &   0.00 &       1 &      0 &       2 &      1\\
  68 &   22.12 &   0.07 & $-$0.98 &   0.15 &    0.94 &   0.14 &    2.57 &   0.21 &  2.240e-17 &     5 &     3 &    0.52 &    0.40 &   0.00 &       5 &      1 &       3 &      1\\
  69 &   24.59 &   1.46 & $-$3.43 &   1.46 &    1.64 &   1.69 &    4.12 &   1.48 &  1.030e-16 &    22 &     4 &    3.12 &    0.40 &   0.50 &     165 &    161 &       0 &     41\\
  70 &   22.95 &   0.24 & $-$2.06 &   0.26 &    1.88 &   0.28 &    2.51 &   0.30 &  2.036e-16 &    44 &     4 &    5.30 &    0.40 &   0.88 &     168 &     91 &       9 &      3\\
  71 &   22.44 &   0.17 & $-$1.07 &   0.23 &    0.72 &   0.36 &    1.71 &   0.40 &  1.871e-16 &    40 &     6 &    2.89 &    0.40 &   0.23 &      70 &     21 &      10 &      3\\
  72 &   22.71 &   0.23 & $-$1.38 &   0.27 &    0.67 &   0.48 &    2.36 &   0.37 &  9.067e-17 &    19 &     4 &    2.03 &    0.40 &   0.00 &      37 &     12 &       8 &      3\\
  73 &   21.69 &   0.04 & $-$0.25 &   0.16 &    1.19 &   0.08 &    2.43 &   0.07 &  1.377e-17 &     3 &     3 &    0.27 &    0.40 &   0.00 &       2 &      1 &       2 &      1\\
\hline
\end{tabular}
Notes to column headers: (2--3) HST F450W (B) magnitude and their errors; (4--9) colours formed from FUV, F450W, F814W and K filters followed by their errors; (10) \ha\ flux in \ecms\ units; (11--12) \ha\ and \hb\  signal-to-noise ratios (SNR); (13) ratio of \ha\ to \hb\ flux; (14--15) Visual extinction in magnitude and error using an equivalent width of 2~\AA\ to correct for the underlying absorption at \hb\ (see text for details); (16--19)  \ha\ and \hb\ emission equivalent widths and their errors in \AA.
\end{table*}

\subsection{Determination of $A_V$}\label{sec:Av}

We use the aperture fluxes in continuum-subtracted \ha\ and \hb\ images to calculate the visual extinction using the Balmer decrement method for photoionized nebulae \citep[see][]{Osterbrock2006}. We used an intrinsic ratio of F(\ha)/F(\hb)=2.87 corresponding to electron temperature and density typical of 10000~K and 100~cm$^{-3}$, respectively, for the ionized nebulae.
The extinction at the \hb\ wavelength is transformed to the visual extinction $A_V$ using the \citet{Cardelli1989} extinction curve. The emission fluxes of Balmer lines, especially the \hb\ line, in \hii\ region spectra could be affected by possible Balmer absorption from evolved populations \citep[e.g.][]{Gonzalez-Delgado2005}. Effect of this on the derived $A_V$ is negligible for regions that belong to the outer ring. However, for rest of the regions, especially those having EW(\hb)$<$10~\AA,
the derived $A_V$ would be overestimated even with this correction. We hence followed the standard procedure \citep[see][]{McCall1985} of adding 2~\AA\ to the emission EW(\hb) and recalculated the \hb\ flux. The correction for the underlying absorption for the \ha\ line makes negligible difference to the derived $A_V$. The errors in fluxes are propagated to obtain errors on $A_V$. 

\begin{figure}
\includegraphics[width=1\columnwidth]{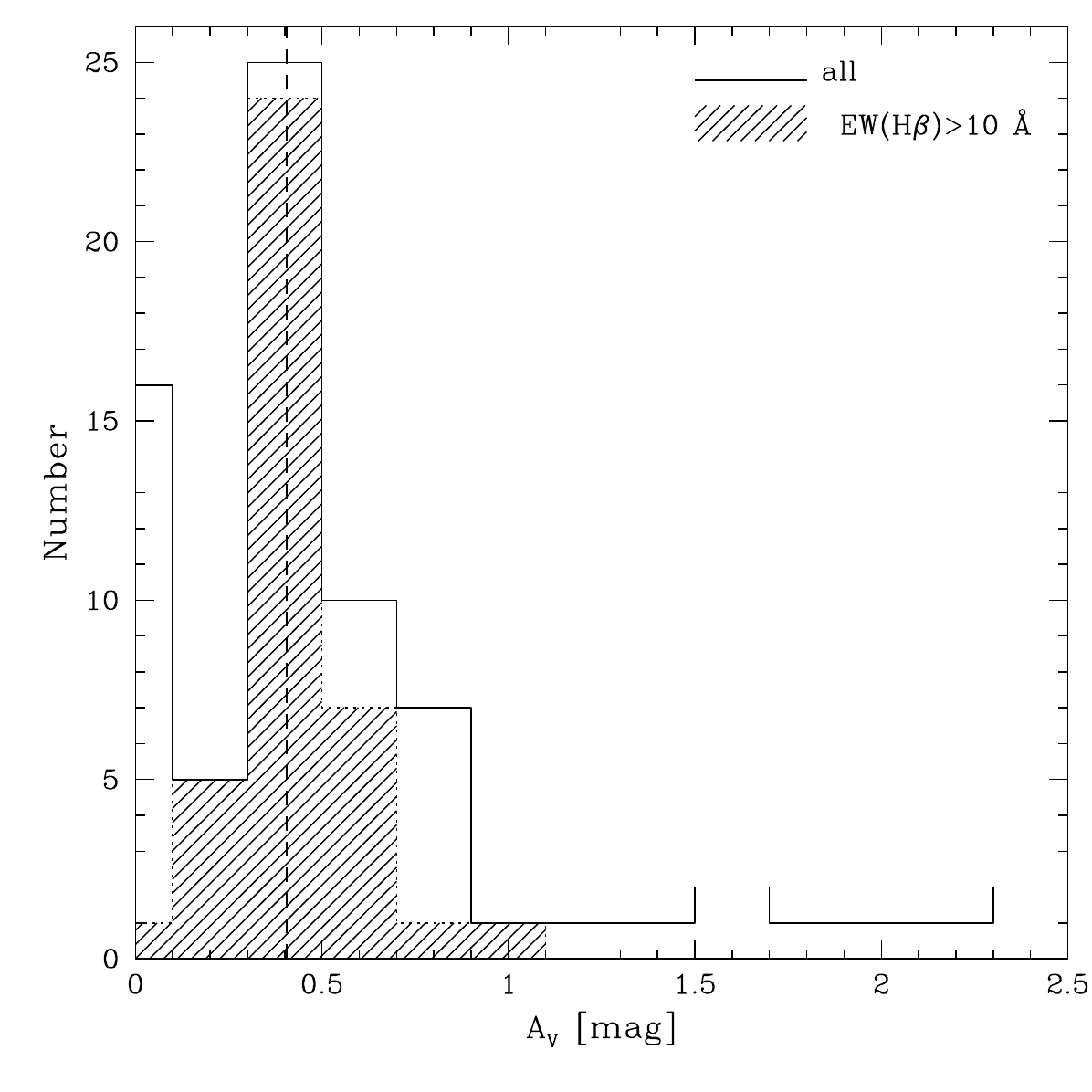}
\caption{
Distribution of $\text{A}_{\text{V}}$ (histogram) derived from Balmer-decrement of the UV-selected sources. The median value of $\text{A}_{\text{V}}=0.4$~mag is marked by a vertical dashed line.
The shaded histogram corresponds to the values for relatively younger regions (\ewhb$>10$~\AA), where the A$_{\rm V}$ could be measured using Balmer decrement method, without getting affected by the underlying stellar absorption features.
}\label{fig:Av}
\end{figure}

The resulting distribution of $\text{A}_{\text{V}}$ values is shown in Figure~\ref{fig:Av}. The distribution is asymmetric with a peak and median value of 0.4~mag, and a mean of 0.54~mag. Most of the ring regions have $\text{A}_{\text{V}}$ values close to the median value, with the tail that extends up to $\text{A}_{\text{V}}=2.5$~mag corresponding to faint regions in the disk
that have EW(\hb)$<10$~\AA. Errors on $\text{A}_{\text{V}}$ ($\delta \text{A}_{\text{V}}$) are large for these regions in the tail and hence are not reliable. We hence reset the calculated $\text{A}_{\text{V}}$ values with the median value for all regions for which EW(\hb)$<10$~\AA.
The resulting values of $\text{A}_{\text{V}}$ are given in column~14 of Table~\ref{tab:obsdata}. Column~15 contains the calculated errors.

\begin{figure*}
\includegraphics[width=1\columnwidth]{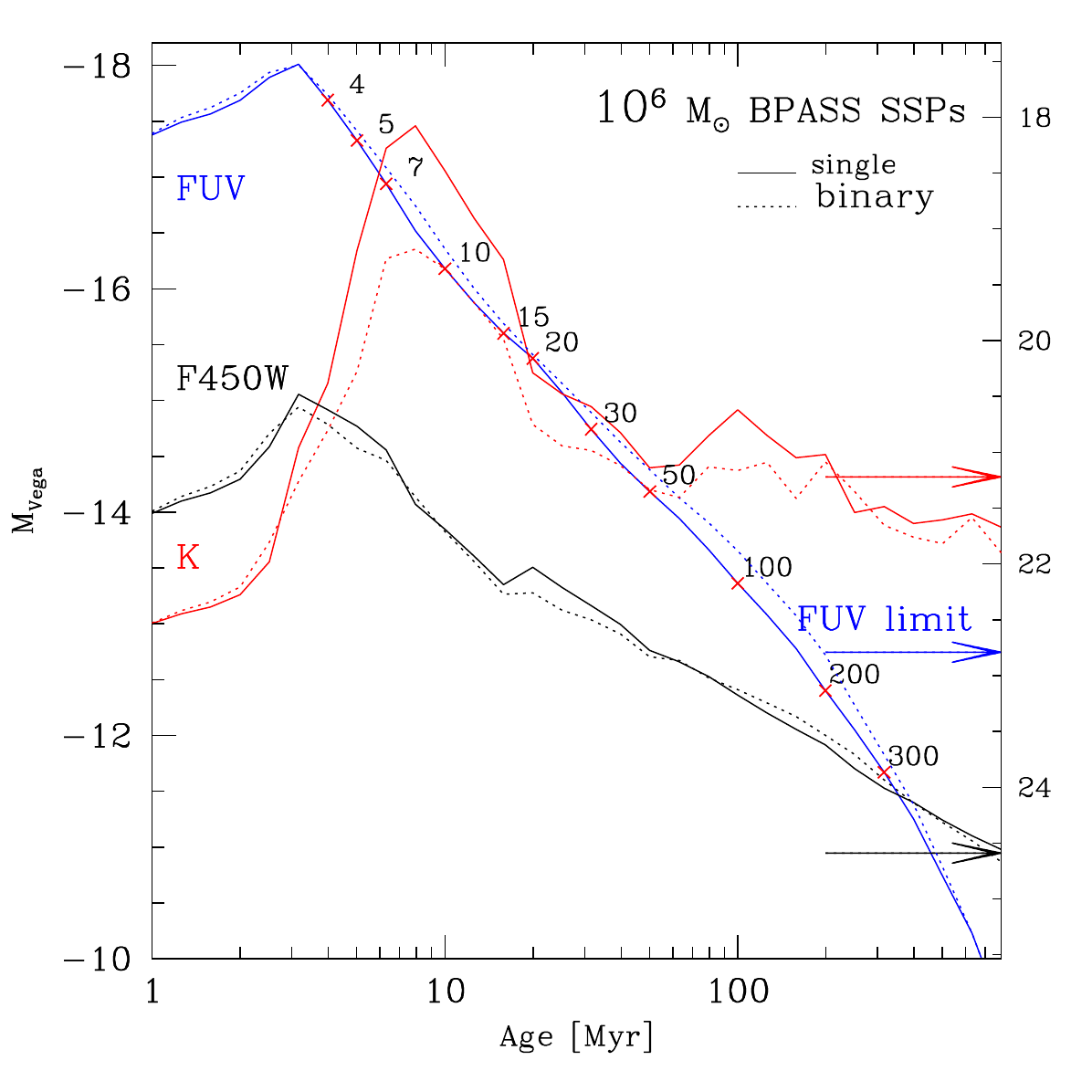}
\includegraphics[width=1\columnwidth]{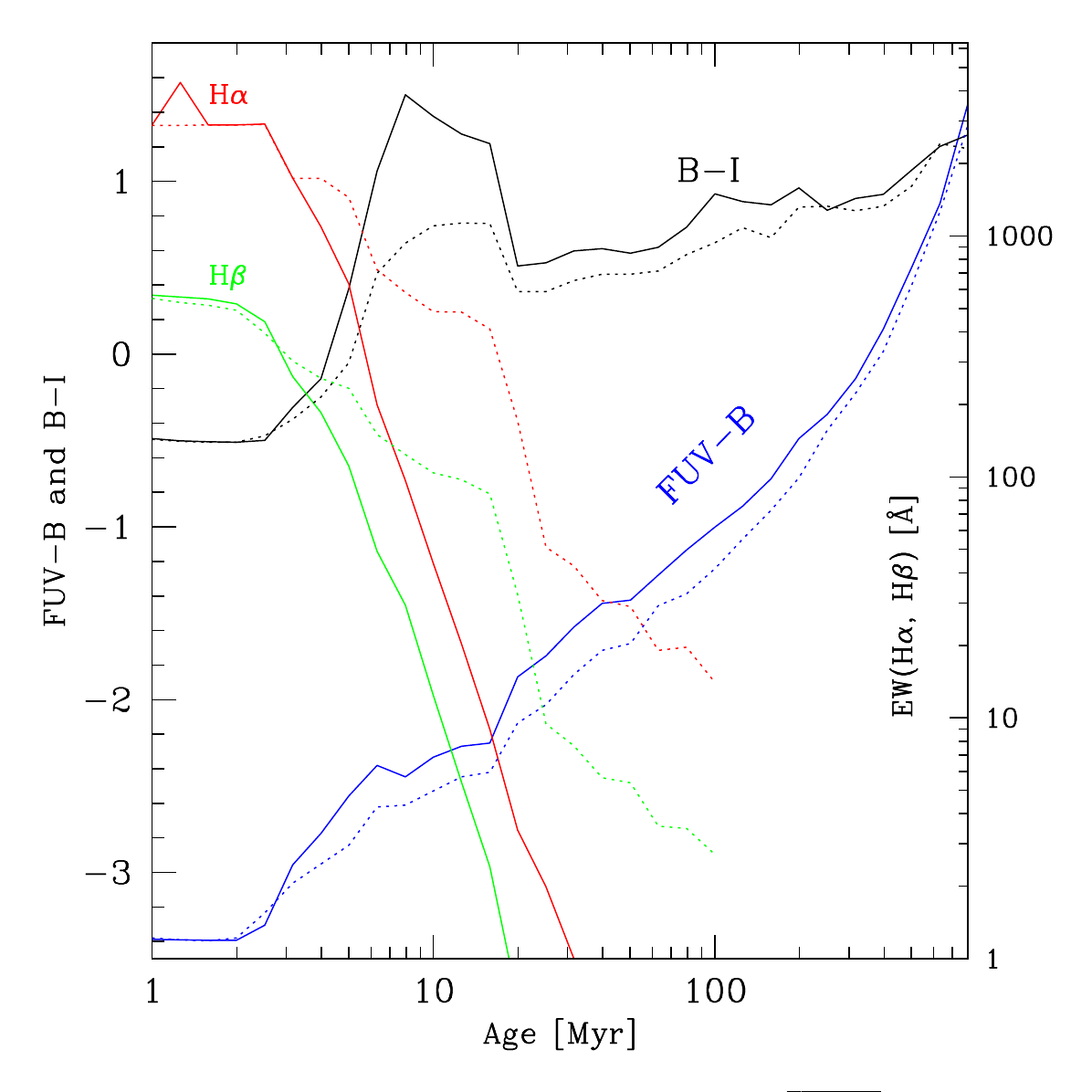}
\caption{
Evolutionary behaviour of quantities used in this work using the BPASS SSP models with (dotted lines) and without (solid lines) binary stars for a metallicity of Z=0.004 and for a stellar mass of $10^6$~\msun. Left panel: absolute magnitudes (left axis scale) in FUV (blue line), Optical (HST F450W; black line) and NIR (K; red line) filters. 
Red crosses 
mark the evolution at selected ages between 4 and 300~Myr. Limiting magnitudes for FUV, F450W and K magnitudes are also indicated by arrows of colours corresponding to the three filters. 
Right panel: FUV-B and B$-$I colours (left axis scale) and \ha\ and \hb\ equivalent widths (right axis scale) are shown by lines of different colours as indicated by the legends.
See text for details. }
\label{fig:model_ssp}
\end{figure*}

\section{Age and mass of compact FUV sources}

\subsection{Evolutionary properties from Population synthesis models}

In Figure~\ref{fig:model_ssp}, we show the evolutionary behaviour of FUV luminosity and colours for 
Simple Stellar Population (SSP)
models using the BPASS (version 2.2) code \citep{Eldridge2017}. Standard Kroupa Initial Mass Function (IMF) is used in these models \citep{Kroupa2001}. We show the plots for Z=0.004 metallicity, which corresponds to that of the Cartwheel \citep[][]{Zaragoza-Cardiel2022}.
BPASS code calculates all physical quantities not only for single star evolution, but also for stellar evolution in binary systems. The left figure illustrates that the UVIT observations are able to detect star-forming complexes as old as $\sim$200~Myr if they have $10^6$~\msun\ of  stellar mass. Older clusters need to be more massive to be detected. On the other hand, clusters of even an order of magnitude less mass can be detected if they are younger than $\sim$10~Myr. All clusters detected in the FUV band can also be detected in the F450W and K-bands. 

The figure to the right shows the evolution of colours formed from the UVIT and HST filters, as well as the equivalent widths of \ha\ and \hb\ nebular lines. The latter quantities, referred to as EW(\ha) and EW(\hb) henceforth, were used from the calculations of  nebular line luminosities for the BPASS models by \citet[][]{Xiao2018}, who carried out the calculations up to an SSP age of 100~Myr. It is worth noting that the evolutionary behaviour of FUV$-$F450W colour is almost monotonous over more than 200~Myr of evolution. On the other hand, in any other colour ($F435W-F814W$ is shown for illustration), the monotonous reddening with age is interrupted by the red peak at around 10~Myr, which correspond to the appearance of Red Super Giants (RSGs) in the SSP models. In the SSP models with binary, the red peak is less pronounced. The absence of the RSG-related bump and a higher slope of the FUV$-$F450W colour compared to any other colour makes the FUV$-$F450W colour an excellent age indicator. 
EW(\ha) and EW(\hb) drop abruptly after the first 3~Myr. The expected values are less than 10~\AA\ at  10~Myr for models without binary stars. The presence of binaries extends the duration for which ionizing photons are available which helps to keep the EW(\ha) and EW(\hb) values marginally above 10 and 3~\AA, respectively, even for systems as old as 100~Myr.
The figure also illustrates that for a given colour,  the age inferred from the binary models is marginally higher than that from single-star models. 

\subsubsection{Dependence of observed quantities on star formation scenarios
}

The plotted models in Figure~\ref{fig:model_ssp} correspond to the evolution of a single Instantaneous Burst  (IB) of star formation, which is a good approximation for individual star clusters. The aperture diameters used for flux measurement in this study correspond to a physical scale of 1.8~kpc in the Cartwheel. The Single-burst scenario is unlikely to be valid over such large scales, especially in a galaxy such as the Cartwheel, which is forming stars for at least the last 80~Myr following the passage of the intruder through the galaxy \citep{Renaud2018}. This interaction is known to have triggered an expanding wave at a velocity of 54~\kms\ \citep{Higdon1996}. At this velocity, the chosen aperture is expected to contain stars formed over the last 32~Myr. The spread of ages could be even larger if clusters formed at inner radii were dragged by the wave as it moved outwards \citep{Renaud2018}. 
Hence, the likely scenario expected in the Cartwheel regions is continuous star formation or the presence of star clusters of multiple ages formed over several tens of million years. The observational quantities plotted in Figure~\ref{fig:model_ssp} vary over different timescales, which allows us to determine the ages of the stellar populations in each of the selected regions. For example, the rapid decrease of emission equivalent widths in IB scenario makes diagrams involving EW(\ha) useful for distinguishing between the single burst, multiple bursts or continuous star formation scenarios. We first make a qualitative comparison of the observed quantities with that expected from different star-formation scenarios in colour-magnitude and colour-EW(\ha) diagrams. These diagrams illustrate that a bursty mode of star formation represents the data better than for a single burst or continuous star formation. We hence adopt a $\chi^2$ fitting to obtain the combination of ages of two representative populations that best reproduce the observed quantities.
\begin{figure*}
\includegraphics[width=1.8\columnwidth]{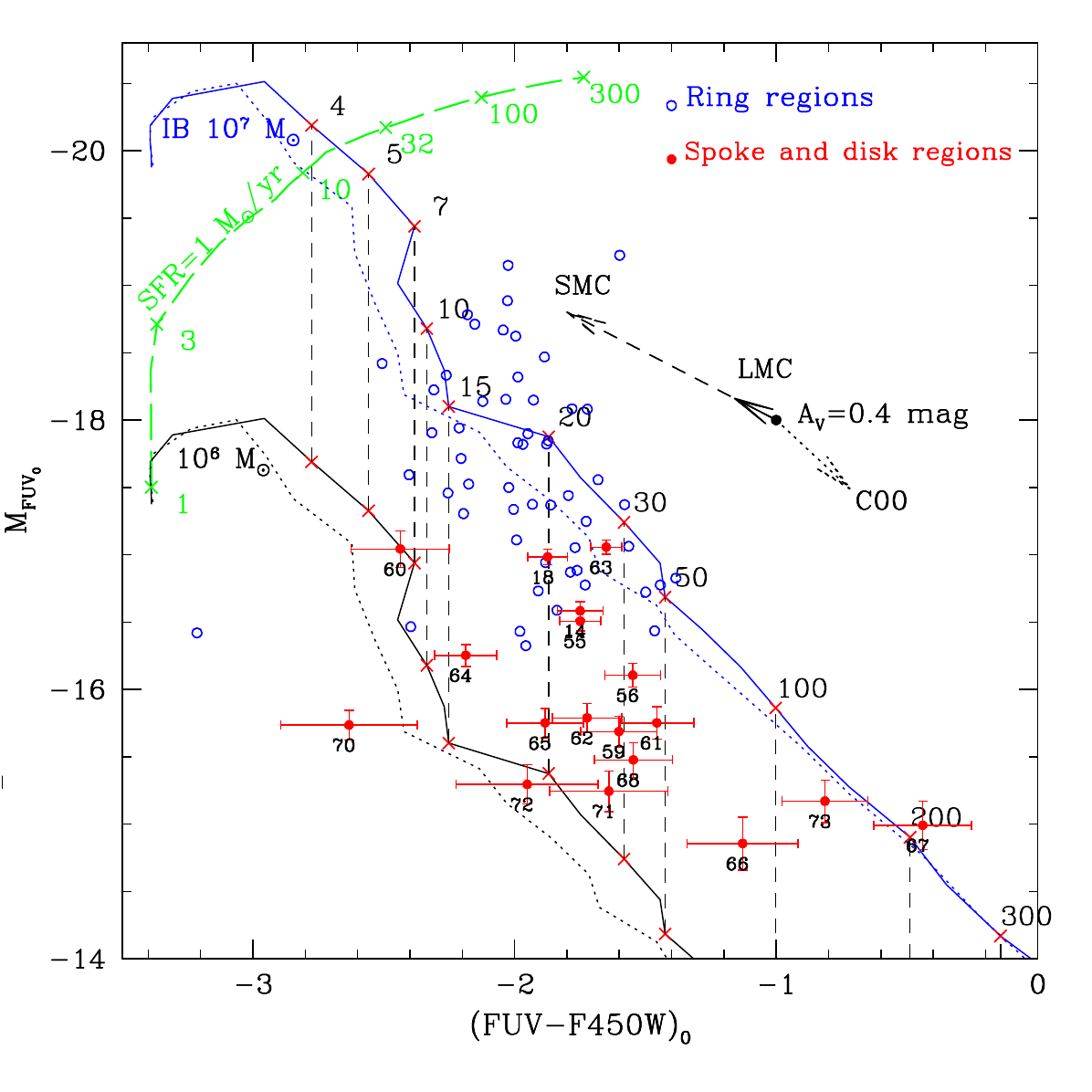}
\caption{
Absolute magnitude ($\text{M}_{\text{FUV}}$) of the FUV sources located in the ring (blue circles) and the spokes or the inner disk (numbered red dots) plotted against their FUV-F450W colours.
Evolutionary track for clusters of masses $10^7$ (blue lines) and $10^6$ \msun\ (black lines) for single and binary (dotted) SSP models are shown,
with the numbers above the cross marks denoting the age in million years. The dashed vertical lines connect the equal-age models for the two plotted cluster masses. The observed magnitudes and colours have been corrected for extinction using A$_{\rm V}$ and \citet{Cardelli1989} extinction curve. Use of the SMC and LMC extinction curves or the \citet{Calzetti2000} attenuation curve would move the observed points by the amount shown by the arrows marked by letters SMC, LMC and C00, respectively, for the mean A$_{\rm V}=$0.4~mag. The green line shows the effect of star formation proceeding continuously for 300~Myr at a constant rate of 1~\Msy, with the tick marks and the associated numbers showing the age in million years. See text for details.
}\label{fig:cmd}
\end{figure*}

\subsection{Effect of using different dust extinction and attenuation curves}

We used the extinction curve of \citet{Cardelli1989} to obtain A$_{V}$ values from the Balmer decrement for the regions studies in this work. The \citet{Cardelli1989} extinction curve is known to represent extinction properties from ultraviolet to near infrared along sight lines to different stars in the Milky Way. \citet{Gordon2003} found that the \citet{Cardelli1989} extinction curve is not appropriate in the ultraviolet wavelengths ($\lambda <$2000~\AA) along majority of sight lines in the Large and Small Magellanic Clouds (LMC and SMC, respectively). The general tendency is for the UV extinction to be higher than the corresponding value from \citet{Cardelli1989}, with the $A_{\rm 1500}/A_{\rm V}$ values reaching values as high as 4.7 for some SMC stars as compared to 2.7 in the Milky Way. Most of the dust causing the extinction to individual stars lies along the line of sight, i.e. in a foreground screen. However, the dust is mixed with the emitting gas in giant star-forming regions, and hence the foreground screen model is a poor approximation when studying extragalactic star forming regions spread over several hundreds of parsecs. \citet{Calzetti1994, Calzetti2000} characterized the attenuation curves over such large scales and found that the $A_{\rm 1500}/A_{\rm V}$ values are lesser than the corresponding MW values in the nuclei of star-forming galaxies. More significantly, even at the optical wavelengths the extinction suffered by the stellar component was found
to be systematically less than that experienced by the nebular gas \citep[see e.g.][]{Calzetti1994, Mayya1996}. \citet{Calzetti2000} established that typically $A_V$(cont) = 0.44$A_V$(gas), where $A_V$(gas) is the visual extinction derived using Balmer decrement values. This is understood to be due to the destruction of obscuring  clouds at the immediate vicinity of the UV and optical continuum-emitting stars by the feedback effects of the massive stars. The nebular flux, on the other hand, originates from  larger scales that contain obscuring dust.

The \citet{Calzetti2000} attenuation curve is expected to be appropriate to the UV-emitting sources in the Cartwheel, given that fluxes are measured using aperture diameters of 1.8~kpc. However, given the large variation in the $A_{\rm 1500}/A_{\rm V}$ values in the LMC and SMC and that the \citet{Cardelli1989} extinction curve represents an intermediate value between the SMC and the extragalactic starburst nuclei, we correct the observed values using the \citet{Cardelli1989} extinction curve and show the effects of extinction curves by vectors in all our diagrams, marked by letters SMC, LMC and C00, which respectively correspond to the average extinction curve from \citet{Gordon2003} for the SMC, 30 Dor star-forming region in the LMC, and the \citet{Calzetti2000} attenuation curve. The plotted vectors show the direction in which the observed points would move if we had used the indicated extinction curve instead of the \citet{Cardelli1989} curve that we have used, with the  arrow sizes corresponding to a nebular extinction of A$_{\rm V}$ = 0.4~mag, which is the average value for our sample. 

\subsection{Colour-magnitude diagram}

In Figure~\ref{fig:cmd}, we plot the photometric properties of the UV selected regions in a colour-magnitude diagram. 
Observed values for regions  belonging to the star-forming outer ring are shown by empty blue circles, whereas those belonging the spoke and disk regions are shown by solid red circles. These latter points are annotated with their identification numbers to help track their location in different figures. The error bars are plotted for the spoke/disk regions. The sizes of the error bars on the ring regions are typically smaller than the spoke/disk regions, and hence we omit their error bars for the sake of clarity of the figure.
BPASS binary evolutionary tracks for SSP populations of $10^7$~\msun\ and $10^6$~\msun\ are shown
in blue and black lines, with the dashed and solid lines corresponding to the SSPs with and without the binary stars, respectively. The location at specific ages are indicated by a red cross on the $10^7$~\msun\ track and are annotated by their ages in million years.

The ring regions are clustered around the track for $10^7$~\msun\ between 10 to 50~Myr age. 
On the other hand, the spoke/disk regions are in general older and less massive. 
The ring regions are associated with bright ionized nebulae that clearly show signs of the presence of massive hot stars in the form of high ionization lines such as He{\sc ii}\,$\lambda$4686 \citep{Mayya2023} and, O{\sc iii}\,$\lambda$5007 \citep[][]{Zaragoza-Cardiel2022}.
Massive hot stars live for less than $\sim$5~Myr, and hence it is surprising that none of the ring regions have colours and UV magnitudes corresponding to SSP ages$<$5~Myr. A different treatment to correct for interstellar dust, and/or the presence of previous generations of stars, are two physical mechanisms that could be the reasons for the observed colours being redder than that expected for an instantaneous burst in its first 5~Myr. We explore these two possibilities below.

The reddening vectors plotted in the Figure suggest that the SMC extinction curve would move the points towards ages $<$5~Myr for majority of the ring regions. On the other hand, the inferred ages would be only marginally younger for the LMC extinction curve, and in fact larger if we had used \citet{Calzetti2000} attenuation curve instead of the \citet{Cardelli1989} extinction curve. Thus, the UV-selected ring regions could be young single-burst populations if the UV extinction curve in the Cartwheel is similar to that for individual stars in the SMC. 

Another possibility suggested above is the presence of relatively older generations of stars in the apertures used for measurements. We recall that in the process of obtaining the aperture magnitudes, the contribution from the underlying disk populations has been subtracted out, and hence if an older generation contributes, it should have been formed in the expanding wave. In Figure~\ref{fig:cmd}, we show in dashed green curve the expected  locus of a continuous star formation model at a constant rate of 1~\msun\,yr$^{-1}$ for up to 300~Myr. The locus will shift downwards for a lower value of star formation rate (SFR) by $\Delta m=2.5\log({\rm SFR})$. The observed colours of ring regions correspond to ongoing star formation for 30--300~Myr at rates of 0.025 to 0.04~\msun\,yr$^{-1}$. 

The colour-magnitude diagram alone cannot distinguish between effect of redddening and the presence of older populations. On the other hand, as we discuss below, plots involving EW(\ha) help to break this degeneracy.

\subsection{EW(H$\alpha$) vs Colour diagram}

\begin{figure*}
\includegraphics[width=1.8\columnwidth]{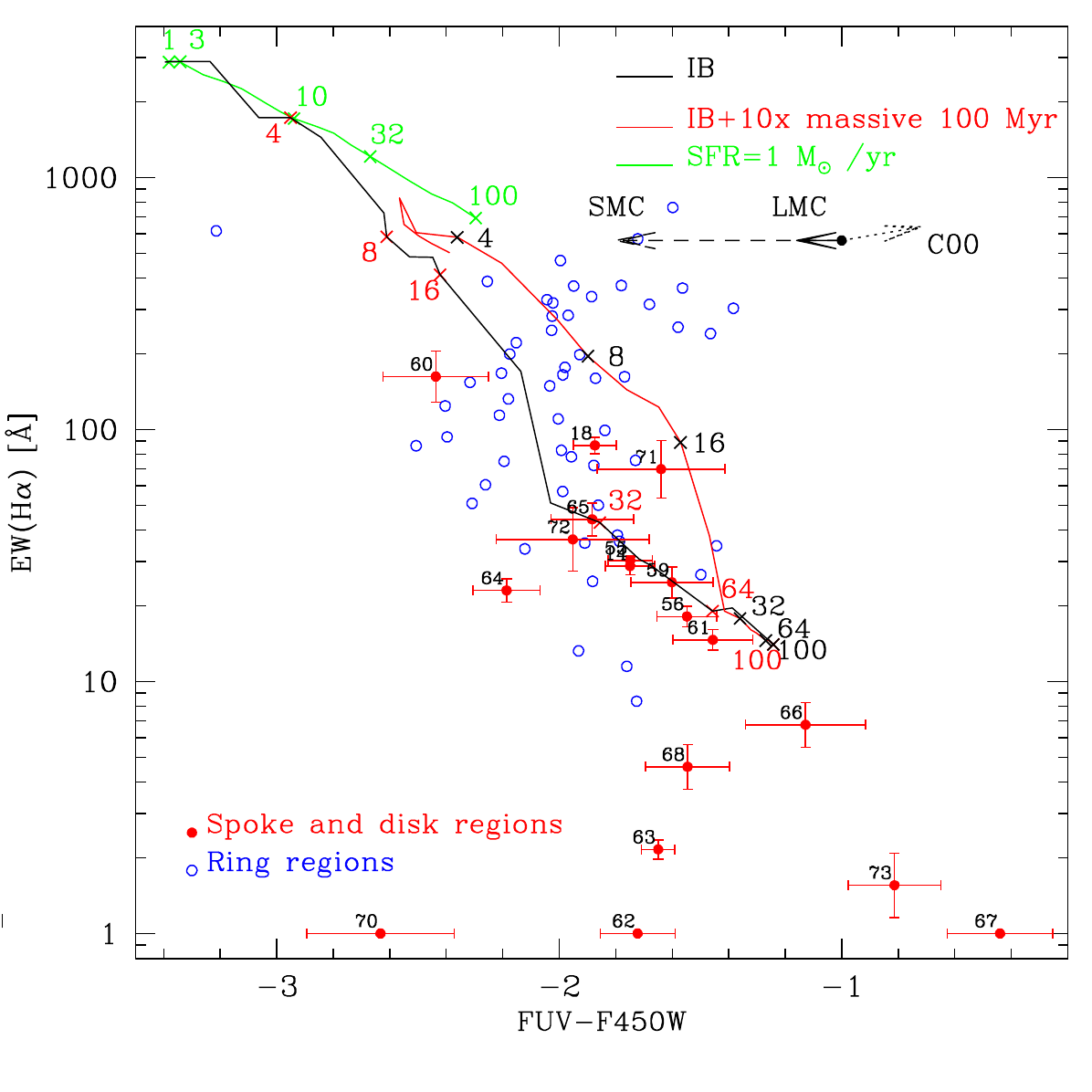}
\caption{
EW(\ha) of the FUV sources located in the ring (blue circles) and the spokes or the inner disk (numbered red dots) plotted against their FUV-F450W colours. The black and green lines show the trajectory for IB and continuous star formation models, respectively, for the BPASS binary tracks. The observed quantities have been corrected for extinction using A$_{\rm V}$ and \citet{Cardelli1989} extinction curve. Use of the SMC and  LMC extinction curves  or the \citet{Calzetti2000} attenuation curve would move the observed points by the amount shown by the arrows marked by letters SMC, LMC and C00, respectively, for the mean A$_{\rm V}=$0.4~mag. Effect of a 
massive ($\times10$) old  burst (100~Myr) superposed on bursts of age between 1 and 100~Myr is shown by the red line. Numbers accompanying the tick marks on the model curves indicate the ages in million years.
}\label{fig:ew_col}
\end{figure*}

In order to understand the reasons 
for the observed FUV$-$F450W colours of ring regions being redder than that expected for populations producing copious amount of ionization,
we plot in Figure~\ref{fig:ew_col}, EW(\ha) against the FUV$-$F450W colour. The observed points and the reddenning vectors follow the same convention as in Figure~\ref{fig:cmd}. The locus of an IB binary model from BPASS \citep{Eldridge2017} is shown in solid black line. It may be noted that the EW being the ratio of emission line to continuous fluxes is independent of the cluster mass. It is also independent of extinction if the obscuring dust is in the foreground such as the case for the LMC, SMC and MW extinction curves. It may have a small dependence on extinction if the absorbing dust is mixed with the gas, and the stars and gas have different spatial extent, as is normally the case in complex star-forming regions \citep{Calzetti2000}. The plot also contains the locus for continuous star formation (green line) and for a model that contains two populations of distinct ages, with the older one contributing negligibly to the ionization (red line). The plotted model corresponds to a 100~Myr old population that is 10 times more massive than the young population of the annotated age. The continuous star formation is carried out for a constant SFR=1~\msun\,yr$^{-1}$. However, the observed locus is independent of the SFR as long as it is constant over the plotted duration of 100~Myr. 

As illustrated in Figure~\ref{fig:model_ssp} (right),   the EW(\ha) for the  IB model falls rapidly for ages greater than $\sim$3~Myr, with its value  being $<$10~\AA, for ages $>$100~Myr. 
The observed EW(\ha) for the ring regions are systematically lower than that expected for populations younger than 16~Myr, with both the observed FUV$-$F450W colour and EW suggesting burst ages between 16 and 64~Myr. 
As noted while discussing the colour-magnitude diagram in the previous section, these ages are too large to explain the high ionization state of the ring regions. 
The CSF model provides the required ionization. However, the EW(\ha) for the CSF model lies above 600~\AA, the highest value observed for our sample, even after star formation proceeding for 100~Myr. Thus the CSF scenario can be clearly ruled out in all observed regions. On the other hand, existence of an old massive non-ionizing population, such as illustrated by the model represented by the red line, is able to reproduce the observed values of EW(\ha) for the young population ages of 4 to 10~Myr. The mean of the observed colours of the ring regions are also consistent with the colour of the combined populations. However, there is a spread of 0.5~mag in colour around the mean value. Variation in extinction properties from region to region could be the reason for this dispersion.

The EWs and FUV$-$F450W colours of the disk and spoke regions (red points) follow the IB evolutionary trajectory until the last available SSP age of 100~Myr, with the majority of points in agreement with the IB scenario between ages of 32 to 64~Myr.
Seven regions (ID numbers 62, 63, 66, 67, 68, 70 and 73) that have the lowest observed EWs, seem to lie around the trajectory that is obtained by extrapolation of the age$<$100~Myr track, suggesting ages larger than 100~Myr.

\subsection{Best-fitting SSP models}

We used a $\chi^2$ minimization procedure to obtain the ages and relative masses of two SSPs whose combined contributions best fit the observed FUV$-$F450W colours and EWs. The $\chi^2$ is defined as:
\begin{equation}
\chi^2 = \left(\frac{C_{\rm obs} - C_{\rm mod}}{\sigma_c}\right)^2 + \left(\frac{lgEW_{\rm obs} - lgEW_{\rm mod}}{\sigma_{\rm ew}}\right)^2,
\label{eqn:chisq}
\end{equation}
where $C_{\rm obs}$ and $C_{\rm mod}$ are the observed and model FUV$-$F450W colours, respectively. Similarly $lgEW_{\rm obs}$ and $lgEW_{\rm mod}$ are the observed and model \ha\ equivalent widths in logarithmic units, respectively. The error terms $\sigma_c$ and $\sigma_{\rm ew}=\log(1+{\rm err}_{ew}/EW)$ are the errors in the observed FUV$-$F450W colours and EWs, respectively. The model colour and EW are defined as the combined values for the two populations, which depend on the mass and age of the two populations. The observed \ha\ luminosity and FUV magnitudes, after extinction corrections  are used to constrain the masses of the two populations. 

As a first step, we determine the mass of the young population using the observed \ha\ luminosity and assuming that the entire observed ionizing flux comes from this young population. With this assumption, the error in the determined mass would be less than 10\% if the second population is older than $\sim$10~Myr. 
\begin{equation}
    \mathcal{M}_{\rm young} = \frac{L(\rm H\alpha)_{\rm obs}}{L(\rm H\alpha)_{\rm young}},
\end{equation}
where ${L(\rm H\alpha)_{\rm young}}$ is the \ha\ luminosity per unit cluster mass for the binary models of BPASS following the nebular line luminosity calculations of \citet[][]{Xiao2018}.
We then calculated the FUV luminosity produced by this population and subtracted this from the observed FUV luminosity to estimate the residual FUV luminosity. We obtained the mass of a second older population that correspond to the residual FUV luminosity, as given by the following equation:
\begin{equation}
    \mathcal{M}_{\rm old} = \frac{10^{-0.4M_{\rm FUV}^0({\rm obs})} - 10^{-0.4M_{\rm FUV}({\rm young})}\times \mathcal{M}_{\rm young}}{10^{-0.4M_{\rm FUV}({\rm old})}}.
\end{equation}

The $C_{\rm mod}$ and $EW_{\rm mod}$ in equation~\ref{eqn:chisq} are calculated for the combined population using the following equations:

\begin{equation}
    C_{\rm mod} = -2.5\log(L_{\rm FUV}({\rm 2pop})) + 2.5\log(L_{\rm F450}({\rm 2pop})) 
\end{equation}

\begin{equation}
    EW_{\rm mod} = \frac{EW({\rm H}\alpha)({\rm young})}{D},
    \label{eqn:dil_fac}
\end{equation}
where $L_{\rm FUV}$(2pop) and $L_{\rm F450}$(2pop) are the sum of the model luminosities in the FUV and F450W filters from the young and old populations, and $D=L_{\rm H\alpha c}({\rm 2pop})/L_{\rm H\alpha c}({\rm young})$ is the dilution factor of EW(\ha) due to the continuum at the \ha\ wavelength from the non-ionizing older population.

The age of the second population is constrained to be older than that of the young population. We calculated  $\chi^2$ by varying the age of the younger population between 1 and 15~Myr and that of the older population between that of the young population and up to 200~Myr in logarithmic steps of 0.1~dex. The age that gives the minimum $\chi^2$ is stored as the course value of age, $t_{\rm course}$. In a second iteration the $C_{\rm mod}$ and $EW_{\rm mod}$ are interpolated to get their values at logarithmic age steps of 0.01~dex between $\log t_{\rm course}-0.1$ and $\log t_{\rm course}+0.1$. The $\chi^2$ minimization procedure is repeated to obtain the new age with this fine interval. The masses of the young and old populations corresponding to the best-fit age are stored.

\begin{figure*}
\includegraphics[width=1.5\columnwidth]{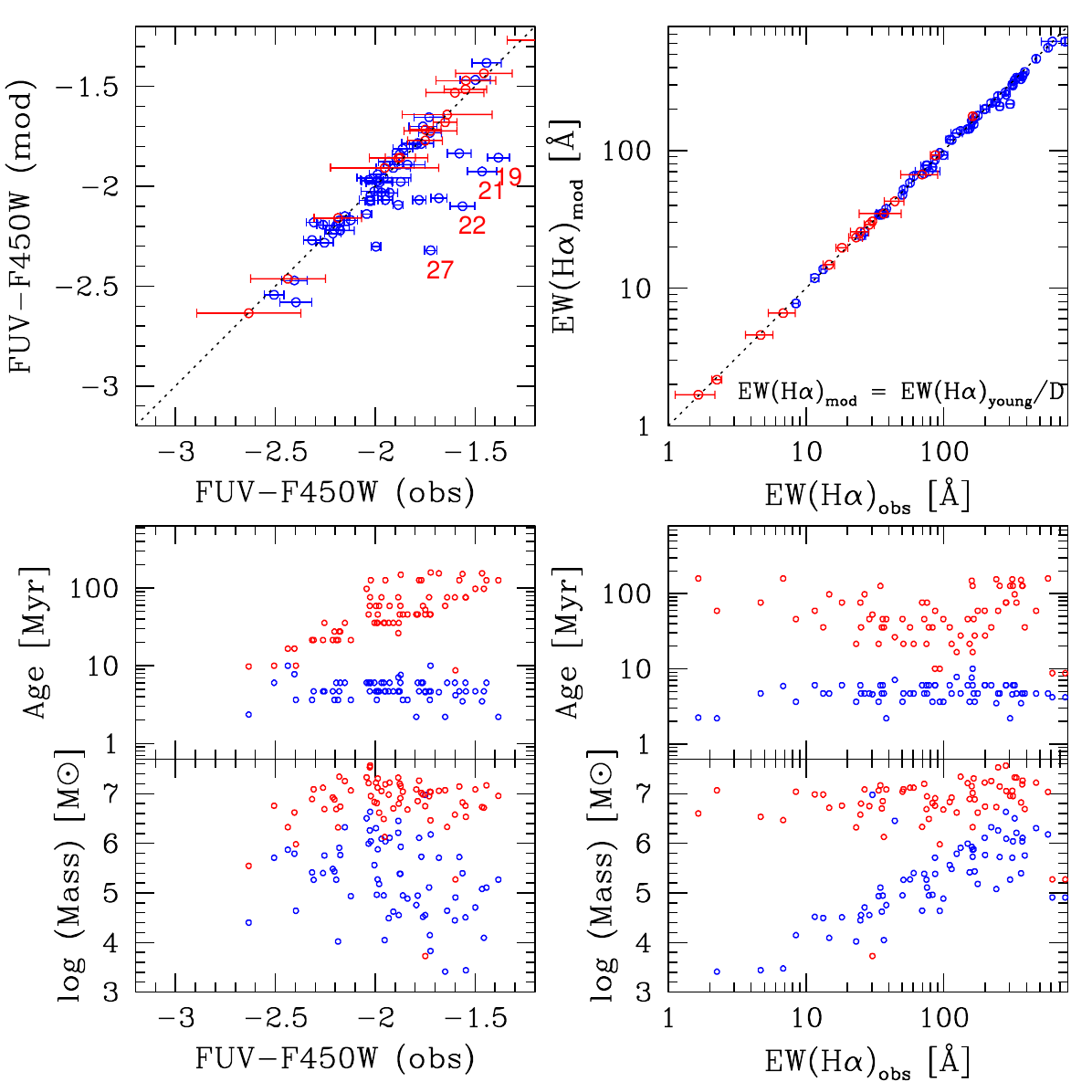}
\caption{
Results of the $\chi^2$-fitting procedure.
(top panels) Model and observed colours FUV--F450W (left) and EW(\ha) (right) are compared with that of the 2-population models;
(bottom panels) individual ages and mass of FUV sources in the ring and spokes of the Cartwheel as a function of observed colour FUV--F450W (left)  and EW(\ha) (right).
See Section~4.6 for details.
}\label{fig:fit_results1}
\end{figure*}

\subsection{Age and mass of FUV sources in the ring and spokes}

The results of the  $\chi^2$-fits are summarized in Figure~\ref{fig:fit_results1}. In the top two panels the observed colours and EWs are compared with that of the 2-population models. Our $\chi^2$-fitting procedure is able to simultaneously reproduce the observed colours and EWs over the entire range of observed values except for a group of $\sim$10 regions with FUV$-$F450W$>-$2.0. Five of these having  extreme differences are labled with their identification numbers.  These are the same regions that lie to the right of the red line in Figure~\ref{fig:ew_col} (EW(\ha)$\sim$400~\AA\ and FUV$-$F450W$\sim-$1.5). Differential reddening between stars and gas following \citet{Calzetti2000} is the most likely reason why our $\chi^2$ fitting  using \citet{Cardelli1989} were not able to reproduce these values.

\begin{figure}
\includegraphics[width=1\columnwidth]{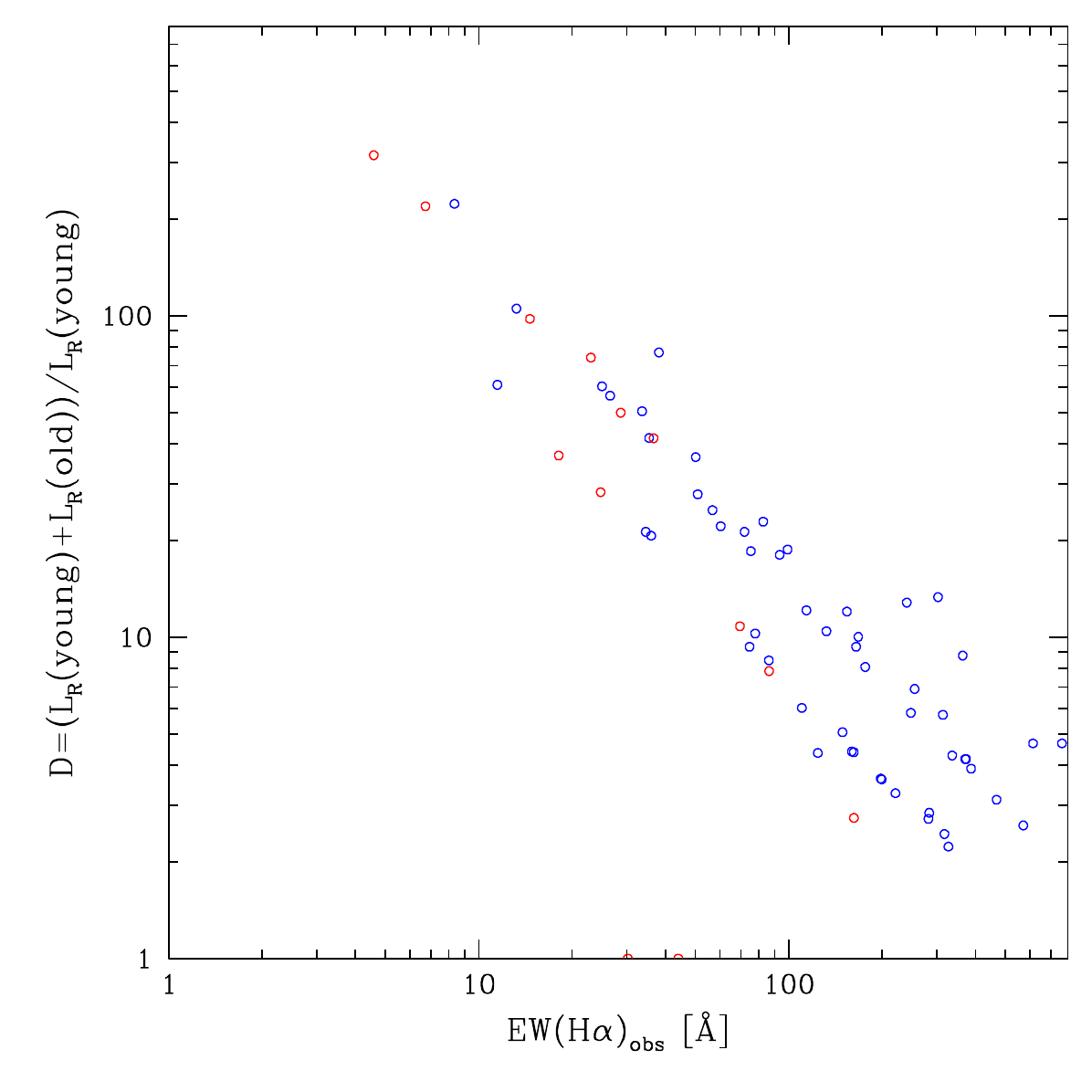}
\caption{
Results of the $\chi^2$-fitting procedure.
Dilution factor D (see equation~\ref{eqn:dil_fac}) vs the observed EW(\ha) of the UV sources in Cartwheel.
}\label{fig:fit_results3}
\end{figure}

In the bottom panels of Figure~\ref{fig:fit_results1}, we show the individual ages and masses of the two populations as a function of observed colours and EWs. The age of the older population is proportional to the observed colour, indicating that the FUV$-$F450W colour has very little contribution from the population that is contributing ionizing photons. This is due to an order of magnitude higher mass of the older population as compared to that of the younger population. The relatively large mass of the non-ionizing population makes the observed EW(\ha) insensitive to the age of the younger population. Instead, the EW is dictated by the mass, or equivalently the \ha\ luminosity of the younger population, i.e. regions luminous in \ha\ have systematically higher EW and vice versa. The presence of a dominant older population makes the observed EW(\ha) dependent strongly on the dilution factor D (see equation~\ref{eqn:dil_fac}), as is illustrated in Figure~\ref{fig:fit_results3}.

\begin{figure}
\includegraphics[width=1\columnwidth]{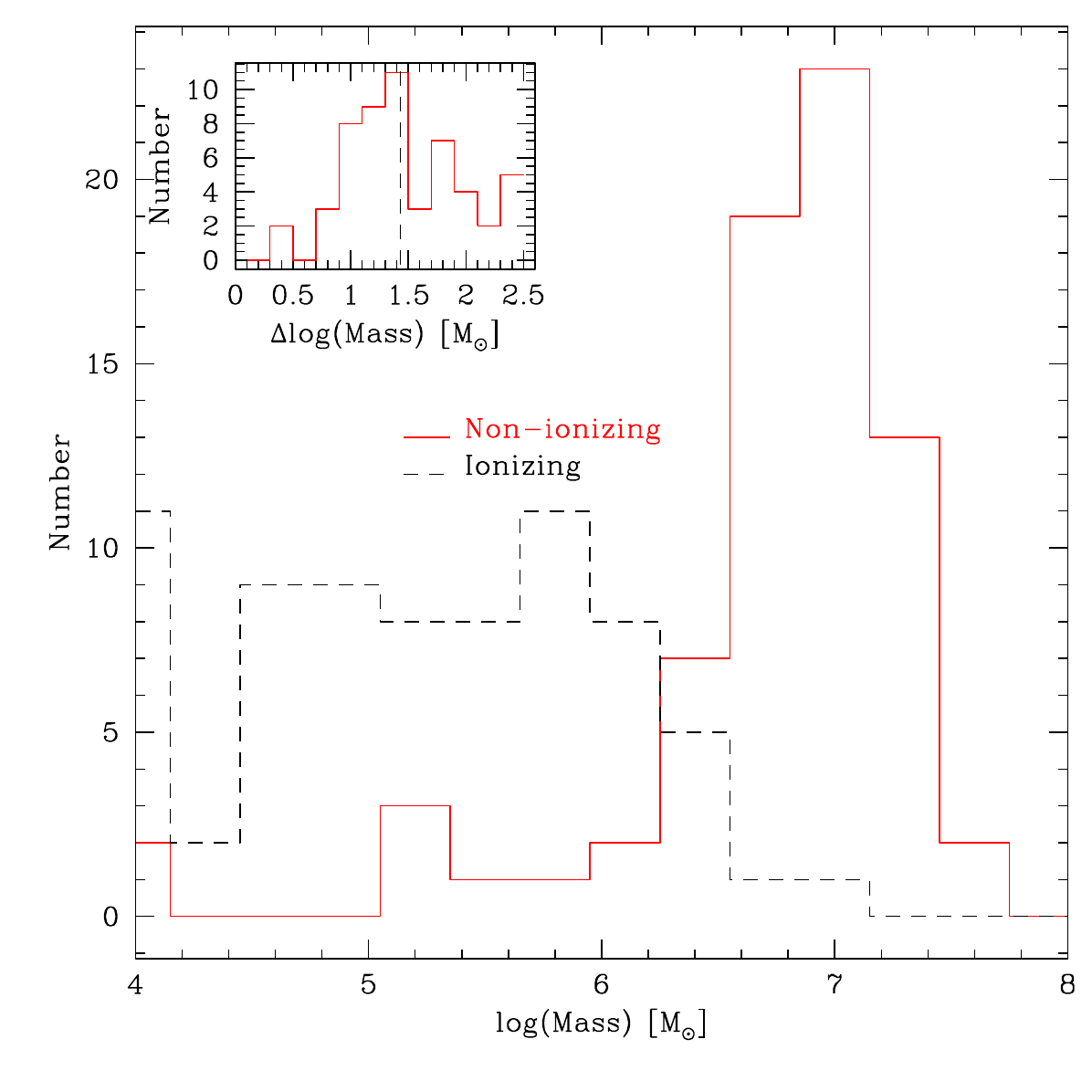}
\caption{
Distribution of masses (histogram) of the non-ionizing (red line) and ionizing (dashed black line) populations in the ring regions of Cartwheel. The non-ionizing population traces relatively older population that has a mean mass of $10^7$~\msun, which is $\sim$25 times larger than the mean mass of the younger population.
}\label{fig:mass}
\end{figure}

In Figure~\ref{fig:mass}, we compare the distribution of masses of the young and old populations. The mean mass of older non-ioning population is $10^7$~\msun, which is $\sim$25 times larger than the mean mass of the younger ionizing population (see the inset).

\section{Discussion}

Having obtained the ages and masses of stellar populations in the outer ring and the disk, including some regions belonging to the spokes, we now analyse the star formation history of the Cartwheel in the context of the collisional model of the formation of the star-forming ring. 

\subsection{Radial colour gradient}

\citet{Marcum1992} found a strong radial colour gradient in Cartwheel which they found to be consistent with a sequential aging of star formation following the ring-making collision.
\citet[][]{Korchagin2001} found a non-negligible contribution from the stars belonging to the pre-collisonal disk to the observed colour. The radial color gradient was inferred by \citet{Marston1995} using azimuthally averaged colours.
The region between the outer and inner rings of the Cartwheel has considerable structures, including the presence of spokes, which makes the azimuthal averaged colours a poor representation of the real variations. We here use our UV-selected sources, without azimuthal averaging, to study the radial gradients in age.

\begin{figure}
\includegraphics[width=1\columnwidth]{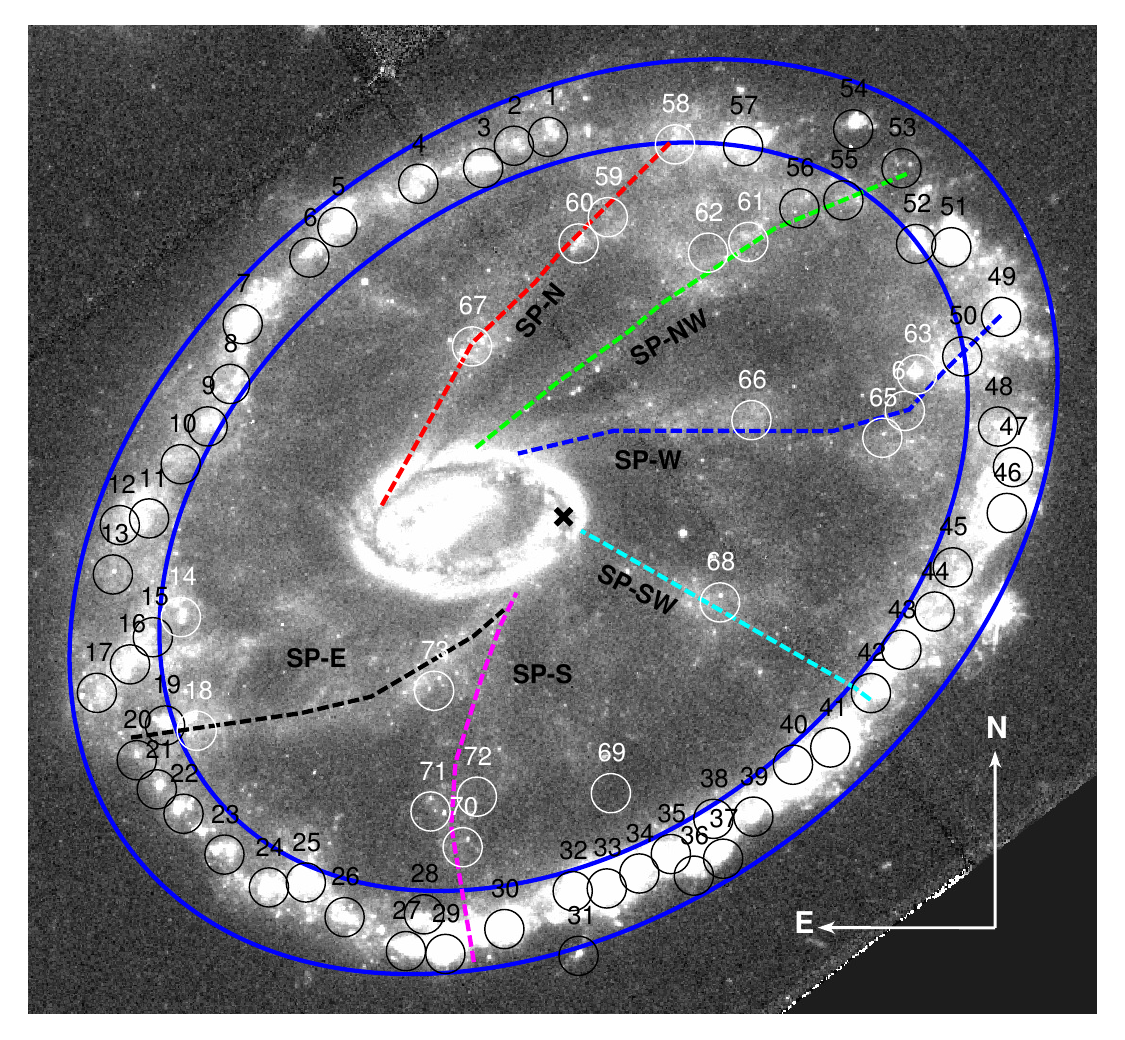}
\caption{
Identification of prominent spokes on the F435W image of the Cartwheel. 
Dashed lines of different colours are drawn to identify each spoke, which are labelled by letters SP-N, SP-NW, SN-W, SP-SW, SP-S and SP-E. Two concentric ellipses delimit the UV sources belonging to the star-forming ring (black circles). The back cross shows the ellipse center.
The majority of the UV selected sources between the inner and outer rings (white circles) belong to one of the six spokes.
}\label{fig:ellipse}
\end{figure}

\begin{figure}
\includegraphics[width=1\columnwidth]{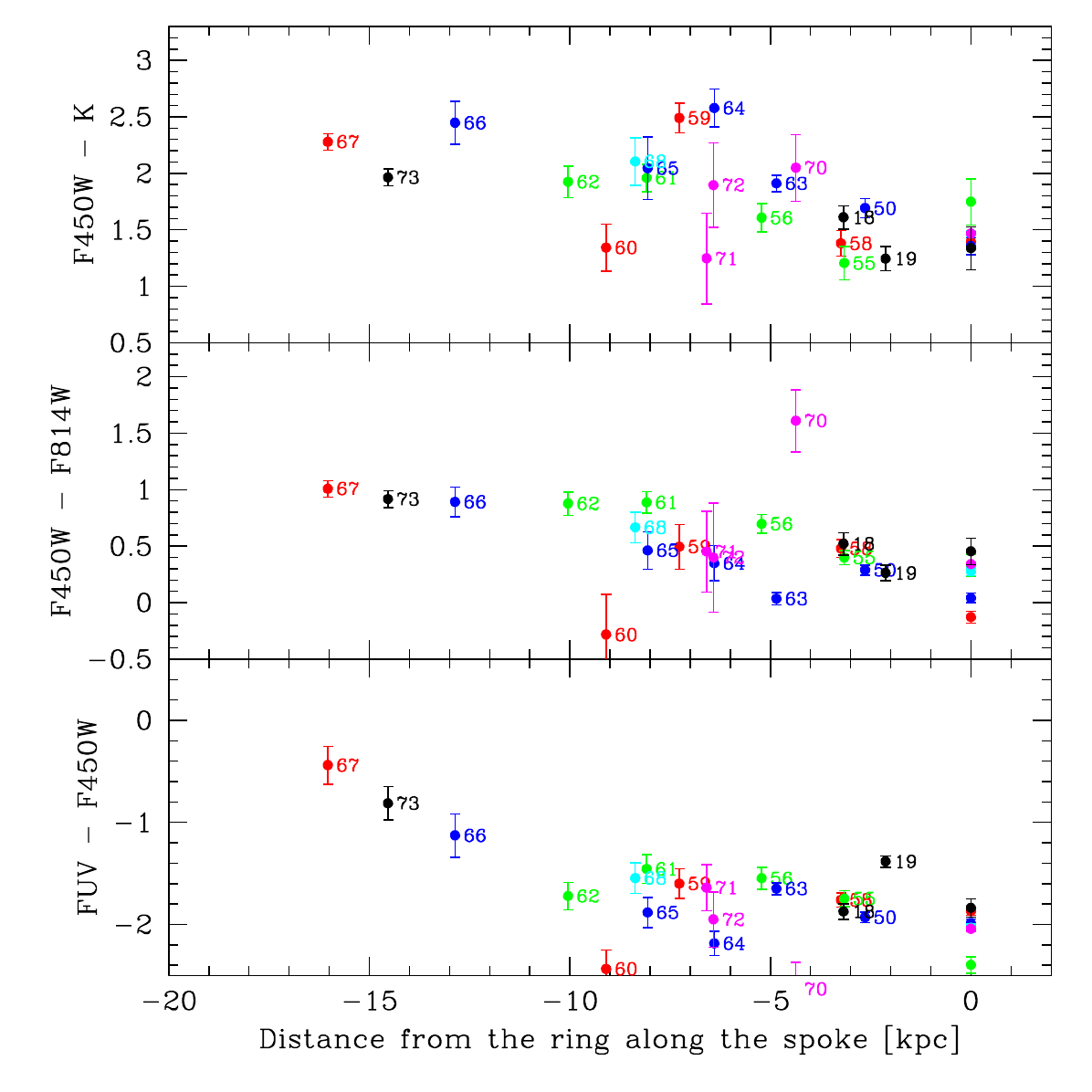}
\caption{Radial variation of the measured quantities, where the distance is measured along the spoke where the region is found.
Regions belonging to the spokes and the region in the ring where the spoke intersects the ring are shown.
All regions belonging to a spoke are identified with the same colour (see Figure~\ref{fig:ellipse}).
All regions in the ring has been assigned a distance of 0 in both the plots. 
}\label{fig:col_gradient}
\end{figure}

In Figure~\ref{fig:ellipse}, we illustrate that the majority of the UV selected sources between the inner and outer rings belong to one of the six spokes traced in this figure, and named clockwise from north by letters SP-N (red), SP-NW (green), SP-W (blue), SP-SW (cyan), SP-S (pink) and SP-S (black). Recent simulation of \citet{Renaud2018} suggest that the spokes might be tracing the trajectory of the material falling back to the nucleus from the outer ring. Based on this suggestion, we measured the distance of a region to the outer ring along the traced spoke. In Figure~\ref{fig:col_gradient}, we show the observed $F450W-K$, $F450W-F814W$ and $FUV-F450W$ colours  plotted against this distance. Only the regions belonging to the spokes and the region in the ring where the spoke intersects the ring, are shown. All regions belonging to a spoke are identified with the 
symbol of the same colour as the colour of the line used in Figure~\ref{fig:ellipse} to delineate the spoke it belongs to. A colour gradient smoothly reddening away from the outer ring is best seen in the FUV$-$F450W colour, with a similar trend in other colours also. The three regions farthest from the outer ring are the reddest. However, in the 10~kpc region internal to the ring, the gradient is shallow. It is worth to note that the range and dispersion in colours among the six points belonging to the ring (xaxis $=0$~kpc) is similar to that of the regions in the 5~kpc zone immediately inside the ring, suggesting that the outer ring, which is easy to delineate in a \ha\ map (see Figure~\ref{fig:ID}, right) will not show up in a colour map. This behaviour illustrates again that the colours are dictated by the non-ionizing populations.

\subsection{Scenario of star formation in the Cartwheel}

The galactocentric distance of any point in the disk of a galaxy is measured as the semi-major axis of the ellipse fitted to the isophote that contains the point. 
The ellipse parameters, namely center, ellipticity and PA are obtained by fitting the isophotes. In spiral galaxies, the ellipticity is a measure of inclination of the disk, which is constant for a non-warped disk, and the center is the nucleus when present, or the center of the bulge, in general. Defining the ellipse center for a collisional ring galaxy is non-trivial as it contains two centers, (1) the nucleus that defines the center of the pre-collisional disk, and (2) the center of the outer ring (the back cross in Figure~\ref{fig:ellipse}). \citet[][]{Marston1995} fixed the nucleus as the center for inner ellipses and varied the center smoothly between the nucleus and the center of the ring to define nine non-intersecting ellipses. The semi-major axis of the ellipse was taken as the galactocentric distance. We followed a similar approach to assign a galactocentric distance to our regions. 

\begin{figure}
\includegraphics[width=1\columnwidth]{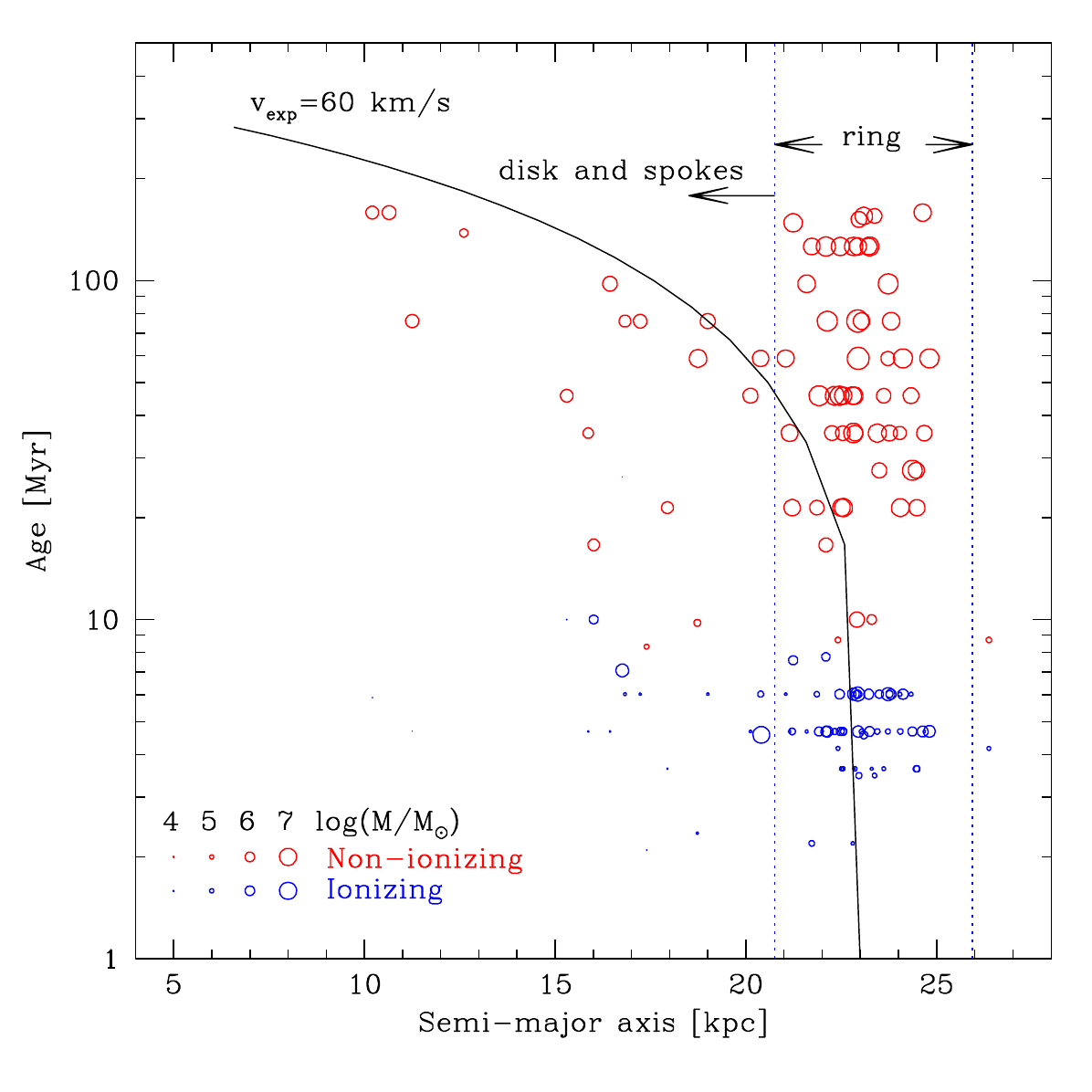}
\caption{
Ages of UV sources are plotted against their current distance from the center of the ellipse that best matches the outer ring. This center is the expected impact point. The ionizing young and non-ionizing old populations of each source are shown by empty circles of blue and red colours, respectively, with the sizes of the circles indicating the mass of the population following size code at the bottom-left corner of the plot. The oldest sources at each distance are consistent with a wave expansion velocity of 60~\kms\ (solid line).
}\label{fig:age_gradient}
\end{figure}

In Figure~\ref{fig:age_gradient}, we plot the derived ages of the young ionizing and old non-ionizing populations in each UV selected region against the galactocentric distance. The size of the symbols is scaled proportional to their masses in order to highlight the most prominent star forming events (see the legends at the bottom-left). The regions belonging to the ring are spread over galactocentric distances between 21 and 26~kpc, which is indicated by vertical blue lines. The age of the old population in the ring covers the entire range of the observed ages. On the other hand, spoke regions farther from the outer ring are systematically older. This trend is especially noticeable in the upper boundary of ages at each galactocentric distance. We show the distance traveled by an outwardly propagating star-forming wave of 60~\kms\ velocity by a solid line, which represents very well the upper boundary of the observed ages as a function of galactocentric distance. We discuss this trend in terms of the collisional scenario of ring formation below.

In the classical scenario of ring formation proposed by \citet[][]{Lynds1976}, the propagating density wave triggers star formation as it advances outwards, leaving behind successively older population of stars in its trail. The age of the oldest population at each radius would then indicate the epoch when the wave propagated at that radius. Once the star formation is triggered it can continue forming new generations of stars until the region runs out of gas or suffers from negative feedback effects of star formation \citep[see e.g.][]{Korchagin2001}. Thus, at each galactocentric distance, a range of younger ages are expected. A comparison of the ages for the older populations in our study with the scenario of star formation triggered by a propagating density wave suggests that the upper boundary of ages at semi-major axis lengths between 10 and 20~kpc is in agreement with a propagating wave of 60~\kms\ velocity, a value consistent with the ring expansion velocity found by \citet{Higdon1996} using HI gas. Thus, the UV-selected regions in the spokes trace star formation that got triggered by the passage of the expanding wave. 

Figure~\ref{fig:age_gradient} shows that the outer ring contains populations formed over a wide range of ages, with the oldest populations as old as those along the spokes. The outer ring is not expected to contain populations older than a few tens of million years according to the classical density wave triggered star formation. This suggests that some of the populations that were formed in the inner parts of the disk during the passage of the density wave were dragged by the wave to their current location in the ring. Such a scenario has been suggested in the recent simulations of the Cartwheel-like galaxies by \citet{Renaud2018}. The relative concentration of UV-selected old regions in the ring and along the spoke suggests that a major fraction of formed stars got dragged in the wave.

\begin{figure}
\includegraphics[width=1\columnwidth]{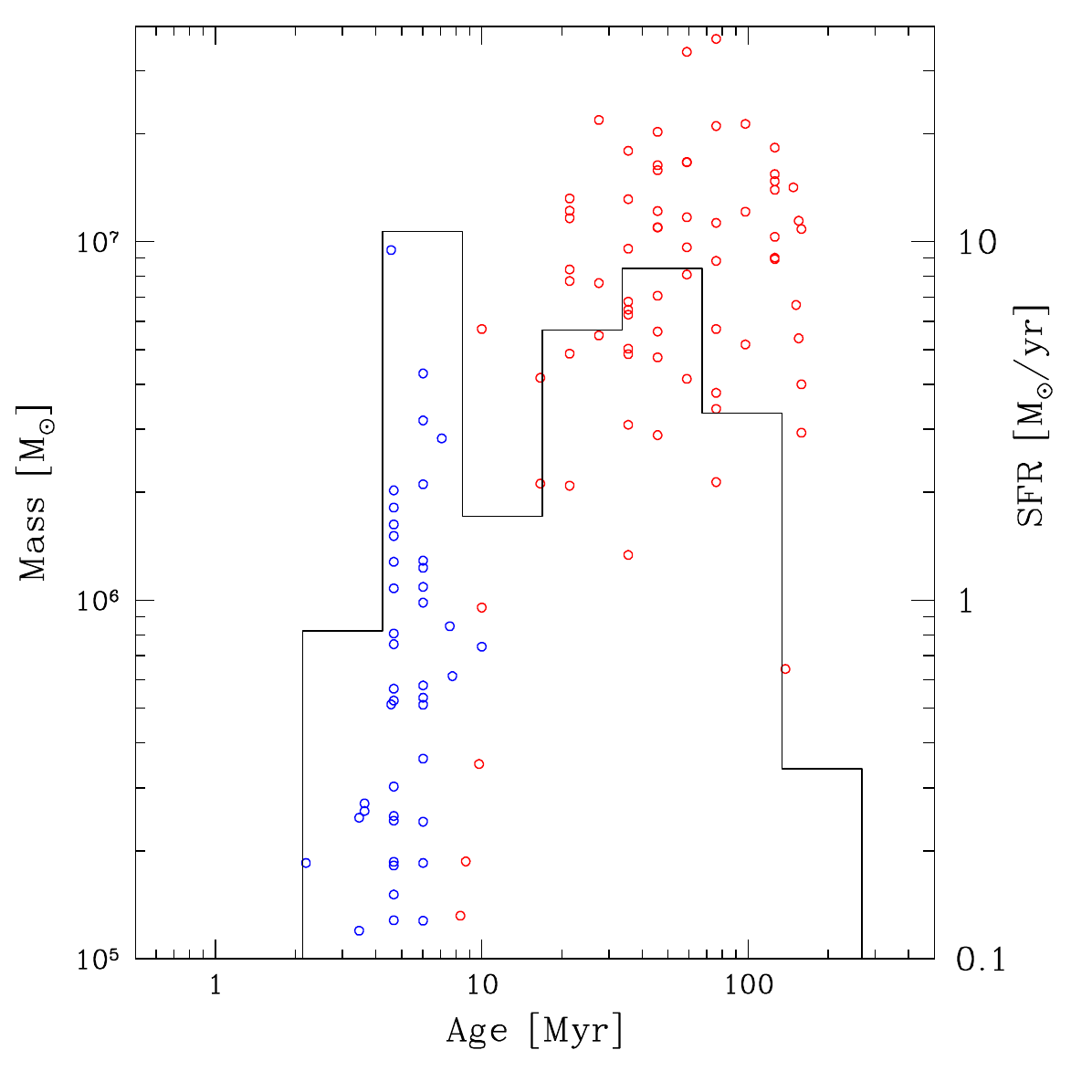}
\caption{
Star formation history of the Cartwheel galaxy using UV data.
The mass (left y-axis) and SFR (right y-axis) of the young (blue circles) and old (red circles) populations are plotted vs the age.
}\label{fig:sfh}
\end{figure}

The availability of mass and age of the old and young star forming events allows us to construct a detailed star forming history of the Cartwheel after it suffered the ring-making collision. In Figure~\ref{fig:sfh}, we show the mass of the young and old populations as a function of their respective ages. We summed the masses of both the populations in bins of 0.5~dex ages to obtain an estimation for the star formation history in the Cartwheel. The results are shown by the solid histogram, with the star formation rate (SFR) indicated in the right axis. At older ages, our detection limit prevents us detecting low-mass clusters, because of which the estimated past SFRs could be lower limits. We find an abrupt increase in star-formation at 150~Myr, which continued at a steady SFR$\sim$5~\msun\,yr$^{-1}$ until $\sim$20~Myr ago. After a quiet period of 10~Myr, star formation became active again over the last 10~Myr, reaching a peak SFR=18~\msun\,yr$^{-1}$, which agrees excellently with the value derived by \citet{Mayya2005} using far infrared, radio continuum and \ha\ luminosities. We remark that the size of the age bins are kept constant in logarithmic intervals, because of which our derived SFR in the past are averages over periods much longer than 20~Myr. Hence, the dip and then a sudden increase in SFR seen over the last 20~Myr may be part of short-term fluctuations. Thus in summary, our analysis suggests that the Cartwheel was forming stars steadily over the last 150~Myr at SFR$\sim$5~\msun\,yr$^{-1}$.

\section{Conclusions}

We investigated the star formation history of the  archetype collisional-ring galaxy Cartwheel.
For this, we analysed new deep and high resolution FUV images obtained with the Astrosat/UVIT mission in the FUV as well as
publicly available optical and IR images. Here we summarize our findings:

\begin{itemize}

\item  The FUV emission of the Cartwheel is concentrated mainly in the star-forming outer ring.
No trace of emission from the nucleus and inner ring is found.

\item  We have identified 73 regions in the Astrosat/UVIT images of the Cartwheel,
58 of which are located in the outer ring. The remaining 15 regions belong to the spokes.

\item  The UV sources in the ring contain more than one population of stars.
The bulk of the FUV emission comes from non-ionizing stars with a range of ages $\sim20-150$~Myr.

\item The FUV regions located in the spokes have negligible current star formation.
The age of the dominant older population increases with the distance from the outer ring.

\item We find that the older populations in the ring have mass that is $\sim$25 times more than the population producing the ionization.

\item The ages for the older populations in our study suggests that the upper boundary of ages at galactocentric distances  between 10 and 20~kpc is in agreement with a propagating wave of 60~\kms\ velocity, consistent with the ring expansion velocity obtained by \citet{Higdon1996}.

\item  We report a SFR $=5$~\Msy over the past 150~Myr, with an increase to $\sim$18~\Msy\ in the recent 10~Myr consistent with \citet{Mayya2005}. 
\end{itemize}

\noindent We conclude that the UV-selected regions in the spokes trace the star formation that got triggered by the passage of the expanding wave. The range of ages of the stellar populations in the ring supports a scenario where some of the stars formed in the wave in the past were dragged along it to the current position of the ring following the scenario suggested by \citet{Renaud2018}. Study of other ring galaxies (e.g. AM\,0644-741) using the Astrosat and optical (e.g. VLT MUSE, HST) data would be able explore whether the scenario presented here for the Cartwheel is applicable for other collisional ring galaxies.
 
\section*{Acknowledgements}
We thank an anonymous referee whose comments have greatly helped in improving the presentation of this paper. 
We also thank CONACyT for the research grant CB-A1-S-25070 (YDM). This publication makes use of data from the Indian Space Research Organization's (ISRO) AstroSat mission, which is housed at the Indian Space Science Data Center (ISSDC). The UVIT data used was processed at IIA by the Payload Operations Centre. The UVIT was developed in collaboration with IIA, IUCAA, TIFR, ISRO, and CSA. This research has made use of the NASA/IPAC Extragalactic Database (NED), which is operated by the Jet Propulsion Laboratory, California Institute of Technology (Caltech) under contract with NASA. 

\section*{Data availability}

Tables~\ref{tab:fuvmag} and \ref{tab:multiband} contain all the astrometric and photometric data used in this work. The reduced fits files on which these data are based will be shared on reasonable request to the first author. The AstroSat UVIT data analysed is publicly available from the ISSDC Astrobrowse archive website: \url{https://astrobrowse.issdc.gov.in/astro_archive/archive/Home.jsp}


\bibliographystyle{mnras}
\bibliography{references}

\end{document}